\documentclass[aps,pra,reprint,showpacs,floatfix,longbibliography]{revtex4-1}
\usepackage{graphicx, amsmath, amssymb, hyperref, multirow}
\usepackage[caption=false]{subfig}
\allowdisplaybreaks[1] 

\begin{document}


\newcommand{\braket}[2]{{\left\langle #1 \middle| #2 \right\rangle}}
\newcommand{\bra}[1]{{\left\langle #1 \right|}}
\newcommand{\ket}[1]{{\left| #1 \right\rangle}}
\newcommand{\ketbra}[2]{{\left| #1 \middle\rangle \middle \langle #2 \right|}}
\newcommand{\fref}[1]{Fig.~\ref{#1}}


\title{Search by Lackadaisical Quantum Walk with Nonhomogeneous Weights}

\author{Mason L.~Rhodes}
	\email{masonrhodes@creighton.edu}
\author{Thomas G.~Wong}
	\email{thomaswong@creighton.edu}
	\affiliation{Department of Physics, Creighton University, 2500 California Plaza, Omaha, NE 68178}

\begin{abstract}
	The lackadaisical quantum walk, which is a quantum walk with a weighted self-loop at each vertex, has been shown to speed up dispersion on the line and improve spatial search on the complete graph and periodic square lattice. In these investigations, each self-loop had the same weight, owing to each graph's vertex-transitivity. In this paper, we propose lackadaisical quantum walks where the self-loops have different weights. We investigate spatial search on the complete bipartite graph, which can be irregular with $N_1$ and $N_2$ vertices in each partite set, and this naturally leads to self-loops in each partite set having different weights $l_1$ and $l_2$, respectively. We analytically prove that for large $N_1$ and $N_2$, if the $k$ marked vertices are confined to, say, the first partite set, then with the typical initial uniform state over the vertices, the success probability is improved from its non-lackadaisical value when $l_1 = kN_2/2N_1$ and $N_2 > (3 - 2\sqrt{2}) N_1$, regardless of $l_2$. When the initial state is stationary under the quantum walk, however, then the success probability is improved when $l_1 = kN_2/2N_1$, now without a constraint on the ratio of $N_1$ and $N_2$, and again independent of $l_2$. Next, when marked vertices lie in both partite sets, then for either initial state, there are many configurations for which the self-loops yield no improvement in quantum search, no matter what weights they take.
\end{abstract}


\maketitle


\section{Introduction}

In a traditional quantum walk \cite{Meyer1996a,FG1998b,Ambainis2001}, a quantum particle scatters or hops to adjacent vertices in superposition, but it has no amplitude of staying put. Such quantum walks have been used to design a variety of quantum algorithms, such as for searching \cite{SKW2003}, solving element distinctness \cite{Ambainis2004}, and evaluating boolean formulas \cite{FGG2008}. Since then, generalizations have been introduced that allow a randomly walking quantum particle to stay put, including lazy quantum walks \cite{Childs2010,Dan2015} and lackadaisical quantum walks \cite{Inui2005,Wong10}. Here, we focus on the lackadaisical quantum walk, where a single, weighted self-loop is added to each vertex \cite{Wong27}. This can cause faster dispersion on the line \cite{Stefanak2012,Wang2017,Wong27}, and it also speeds up quantum search on the complete graph \cite{Wong10,Wong27} and two-dimensional periodic square lattice (or discrete torus) \cite{Wong28,Nahimovs2019}.

\begin{figure}
\begin{center}
	\includegraphics{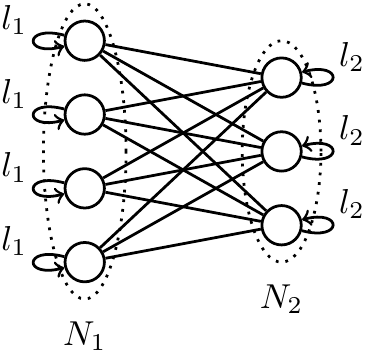}
	\caption{\label{fig:bipartite_unmarked}A complete bipartite graph with $N_1$ vertices in set $X$ and $N_2$ vertices in set $Y$. Each vertex in $X$ ($Y$) contains a self-loop of weight $l_1$ ($l_2)$.}
\end{center}
\end{figure}

In all of these works, each self-loop had the same weight. This is reasonable because the line, complete graph, and periodic square lattice are each vertex-transitive, meaning every vertex has the same structure. Then it is natural for the self-loops to have the same weight, and this preserves the vertex-transitivity of the graph. In this paper, we propose lackadaisical quantum walks where the self-loops can have different weights. Such walks naturally arise when the graph is \emph{not} vertex-transitive, since different vertices have different structures. For example, real-world networks typically have degree distributions that follow power laws \cite{Barabasi1999}, so they contain fewer highly connected hubs and many weakly connected nodes. The different structures of these vertices naturally give rise to self-loops of different weights. As a first step toward future work on complex networks, this paper focuses on one of the simplest graphs for which self-loops of different weights naturally arise: the complete bipartite graph, which is generally irregular and, hence, generally \emph{not} vertex-transitive. An example is shown in \fref{fig:bipartite_unmarked}, where the partite sets $X$ and $Y$ have $N_1$ and $N_2$ vertices, respectively. Since the vertices in each partite set have a different structure, this naturally leads to a lackadaisical quantum walk where the self-loops in one partite set can have different weights from those in the other partite set, which in \fref{fig:bipartite_unmarked} are labeled $l_1$ and $l_2$.

The complete bipartite graph is of interest for several reasons. As discussed above, it is one of the simplest graphs that is generally \emph{not} vertex-transitive, naturally supporting a lackadaisical quantum walk with nonhomogeneous weights. Next, it is highly symmetric, which makes it more amenable to analytical proof, which elucidates the properties that cause speedups or slowdowns. As such, when a new quantum walk is proposed, it is often soon analyzed on highly symmetric graphs \cite{Reitzner2009}. Next, Grover's algorithm is akin to searching the complete graph, or all-to-all network. As the graph loses edges, eventually the Grover speedup is lost. The complete bipartite graph, can have significantly fewer vertices than the complete graph, yet it still supports optimal quantum search in $O(\sqrt{N})$ time \cite{Reitzner2009,Wong31}. Finally, complete bipartite graphs that are additionally regular are examples of strongly regular graphs \cite{Wong5}, which are of significant interest in graph theory.

In this paper, we explore spatial search on the complete bipartite graph using the discrete-time, coined quantum walk \cite{Meyer1996a}, where a particle has a location and direction \cite{Ambainis2001}. We denote a particle at vertex $u$ pointing to vertex $v$ as $\ket{u} \otimes \ket{v} = \ket{uv}$. The walk evolves by $U_\text{walk} = SC$, where $C$ is the Grover diffusion coin generalized to weighted graphs \cite{Wong27}, and $S$ is the flip-flop shift \cite{AKR2005} that causes a particle to hop and turn around. That is, at vertex $u$ with self-loop weight $l$, the generalized Grover coin acts by
\[ C = \ketbra{u}{u} \otimes \left( 2 \ketbra{s_u}{s_u} - I \right), \]
where
\[ \ket{s_u} = \frac{1}{\sqrt{\text{deg}(u)}} \left( \sum_{\substack{v \sim u \\ v \ne u}} \ket{v} + \sqrt{l} \ket{u} \right), \]
where $\text{deg}(u)$ is the degree of $u$, which equals its number of unweighted edges plus $l$ (the weight of the loop). The flip-flop shift transforms $S \ket{uv} = \ket{vu}$, so a particle at vertex $u$ pointing to vertex $v$ hops to $v$ and points toward $u$. Two steps of $U_\text{walk} = SC$ is equivalent to one step of Szegedy's quantum walk \cite{Szegedy2004,Wong27}, which is a quantum process on general Markov chains. To search, we additionally query an oracle $Q$ that flips the signs of the amplitudes at marked vertices. That is, $Q\ket{uv} = -\ket{uv}$ if $u$ is a marked vertex, and $Q\ket{uv} = \ket{uv}$ if $u$ is unmarked. Altogether, the search algorithm repeatedly applies $U = SCQ$.

Following \cite{Wong31}, we consider two different initial states for the search algorithm. The first is the typical uniform superposition over the vertices, where the initial probability of finding the particle at each vertex is $1/N$. At each vertex, the amplitude is distributed among its edges according to weight \cite{Wong27}, and the initial state is:
\begin{align}
	\ket{s} 
		&= \frac{1}{\sqrt{N_1+N_2}} \nonumber \\
		&\quad\times \Bigg[ \sum_{v \in X} \ket{v} \otimes \frac{1}{\sqrt{N_2 + l_1}} \left( \sum_{u \in Y} \ket{u} + \sqrt{l_1} \ket{v} \right) \label{eq:s} \\
		&\quad\quad+ \sum_{v \in Y} \ket{v} \otimes \frac{1}{\sqrt{N_1 + l_2}} \left( \sum_{u \in X} \ket{u} + \sqrt{l_2} \ket{v} \right) \Bigg]. \nonumber
\end{align}
If the graph is irregular, then $U_\text{walk} \ket{s} \ne \ket{s}$, so the system evolves even though we have not obtained any new information about the locations of the marked vertices. So the second initial state $\ket{\sigma}$ is chosen to be stationary under the walk, i.e., $U_\text{walk} \ket{\sigma} = \ket{\sigma}$. The state satisfying this is
\begin{align}
	\ket{\sigma} 
		&= \frac{1}{\sqrt{2N_1N_2 + l_1N_1 + l_2N_2}} \nonumber \\
		&\quad\times \Bigg[ \sum_{v \in X} \ket{v} \otimes \left( \sum_{u \in Y} \ket{u} + \sqrt{l_1} \ket{v} \right) \label{eq:sigma} \\
		&\quad\quad+ \sum_{v \in Y} \ket{v} \otimes \left( \sum_{u \in X} \ket{u} + \sqrt{l_2} \ket{v} \right) \Bigg]. \nonumber
\end{align}

In the next section, we analyze search when the marked vertices lie in a single partite set. Then the system evolves in a 7-dimensional (7D) subspace, permitting asymptotic analysis using degenerate perturbation theory. We prove that for large $N_1$ and $N_2$, then when $l_1 = kN_2/2N_1$, the success probability when searching from $\ket{s}$ changes from its loopless value of $\max(N_1,N_2)/(N_1+N_2)$ to at least $(\sqrt{N_1}+\sqrt{N_2})^2/2(N_1+N_2)$, for any value of $l_2$ (small compared to $N_1$ and $N_2$). This is an improvement when $N_2 > (3 - 2\sqrt{2})N_1$. When searching from $\ket{\sigma}$, the success probability is boosted from $1/2$ to $1$, again independent of $l_2$. This doubling of the success probability occurs regardless of the ratio of $N_1$ to $N_2$. Then in Section 3, the marked vertices can lie in both sets, and the system evolves in a larger 12D subspace. When the search problem is fully symmetric, it can be analyzed, and we prove that self-loops provide no improvement. When the search problem is not fully symmetric, the larger dimensionality is a barrier to analytical proof, so we resort to numerical simulations. Perhaps surprisingly, the self-loops only provide an improvement with initial state $\ket{\sigma}$ when the marked vertices are more dense in one vertex set, indicating that lackadaisical quantum walks are not panaceas for improving all quantum search problems.


\section{Marked Vertices in One Set}

\begin{figure}
\begin{center}
	\includegraphics{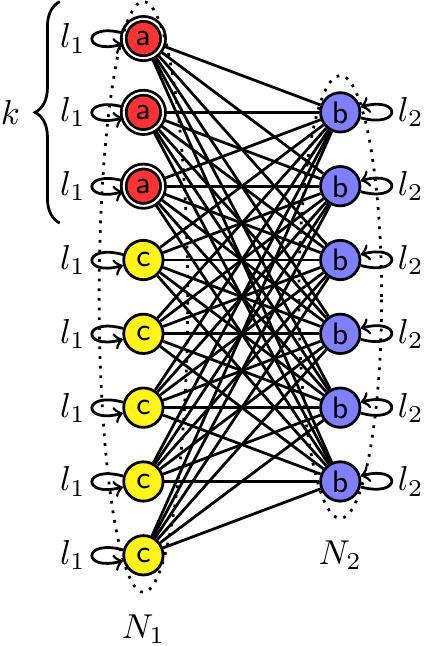}
	\caption{\label{fig:bipartite_onemarked}The complete bipartite graph containing $N_1$ vertices in set $X$, each with a self-loop of weight $l_1$, and $N_2$ vertices in set $Y$, each with a self-loop of weight $l_2$. There are $k$ marked vertices in set $X$, indicated by double circles. Vertices that evolve identically share the same color and letter, and the letters correspond to the states of the subspace basis vectors.}
\end{center}
\end{figure}

Throughout this section, we assume there are $k$ marked vertices in a single partite set, and without loss of generality, say they are in set $X$. This is depicted in \fref{fig:bipartite_onemarked}, and with either initial state $\ket{s}$ or $\ket{\sigma}$, the system evolves in a 7D subspace spanned by the following orthonormal basis states:
\begin{align*}
	& \ket{aa}=\frac{1}{\sqrt{k}}\sum_a \ket{a} \otimes \ket{a}, \\
	& \ket{ab}=\frac{1}{\sqrt{k}}\sum_a \ket{a} \otimes \frac{1}{\sqrt{N_2}}\sum_b \ket{b}, \\
	& \ket{ba}=\frac{1}{\sqrt{N_2}}\sum_b \ket{b} \otimes \frac{1}{\sqrt{k}}\sum_a \ket{a}, \\
	& \ket{bb}=\frac{1}{\sqrt{N_2}}\sum_b \ket{b} \otimes \ket{b}, \\
	& \ket{bc}=\frac{1}{\sqrt{N_2}}\sum_b \ket{b} \otimes \frac{1}{\sqrt{N_1-k}}\sum_c \ket{c}, \\ 
	& \ket{cb}=\frac{1}{\sqrt{N_1-k}}\sum_c \ket{c} \otimes \frac{1}{\sqrt{N_2}}\sum_b \ket{b}, \\
	& \ket{cc}=\frac{1}{\sqrt{N_1-k}}\sum_c \ket{c} \otimes \ket{c}.
\end{align*}
Using (9) from \cite{Wong18}, in this basis, the search operator $U = SCQ$ is
\begin{widetext}
\begin{equation}
	\label{eq:onemarked_U}
	U = \begin{pmatrix}
		\frac{N_2-l_1}{N_2+l_1} & \frac{-2\sqrt{N_2l_1}}{N_2+l_1} & 0 & 0 & 0 & 0 & 0 \\
		0 & 0 & \frac{2k-N_1-l_2}{N_1+l_2} & \frac{2\sqrt{l_2k}}{N_1+l_2} & \frac{2\sqrt{k(N_1-k)}}{N_1+l_2} & 0 & 0 \\
		\frac{-2\sqrt{N_2l_1}}{N_2+l_1} & \frac{l_1-N_2}{N_2+l_1} & 0 & 0 & 0 & 0 & 0 \\
		0 & 0 & \frac{2\sqrt{l_2k}}{N_1+l_2} & \frac{l_2-N_1}{N_1+l_2} & \frac{2\sqrt{l_2(N_1-k)}}{N_1+l_2} & 0 & 0 \\
		0 & 0 & 0 & 0 & 0 & \frac{N_2-l_1}{N_2+l_1} & \frac{2\sqrt{N_2l_1}}{N_2+l_1} \\
		0 & 0 & \frac{2\sqrt{k(N_1-k)}}{N_1+l_2} & \frac{2\sqrt{l_2(N_1-k)}}{N_1+l_2} & \frac{N_1-2k-l_2}{N_1+l_2} & 0 & 0 \\
		0 & 0 & 0 & 0 & 0 & \frac{2\sqrt{N_2l_1}}{N_2+l_1} & \frac{l_1-N_2}{N_2+l_1} \\
	\end{pmatrix}.
\end{equation}

The evolution of the algorithm can be determined by writing the initial state as a linear combination of the eigenvectors of $U$. Then, each application of $U$ multiplies each eigenvector by its eigenvalue. The exact eigenvectors of $U$ are complicated, however, making analysis prohibitive. Yet, they can be approximated for large $N_1$ and $N_2$ using degenerate perturbation theory. The derivation is given in Appendix~\ref{sec:eigen_onemarked}, and the asymptotic eigenvectors and eigenvalues of $U$ are
\begin{align}
	& \ket{\psi_1}=\frac{1}{\sqrt{1+\frac{2l_1 N_1}{k N_2}}}\left[1,0,0,0,-\sqrt{\frac{l_1 N_1}{k N_2}},-\sqrt{\frac{l_1 N_1}{k N_2}},0\right]^\intercal, \> \lambda_1=1, \nonumber \\
	& \ket{\psi_2}=\frac{1}{\sqrt{2+\frac{k N_2}{l_1 N_1}}}\left[1,i\sqrt{\frac{2l_1 N_1+k N_2}{4l_1 N_1}},-i\sqrt{\frac{2l_1 N_1+k N_2}{4l_1 N_1}},0,\sqrt{\frac{k N_2}{4l_1 N_1}},\sqrt{\frac{k N_2}{4l_1 N_1}},0\right]^\intercal, \> \lambda_2=e^{-i\theta}, \nonumber \\
	& \ket{\psi_3}=\frac{1}{\sqrt{2+\frac{k N_2}{l_1 N_1}}}\left[1,-i\sqrt{\frac{2l_1 N_1+k N_2}{4l_1 N_1}},i\sqrt{\frac{2l_1 N_1+k N_2}{4l_1 N_1}},0,\sqrt{\frac{k N_2}{4l_1 N_1}},\sqrt{\frac{k N_2}{4l_1 N_1}},0\right]^\intercal, \> \lambda_3=e^{i\theta}, \nonumber \\
	& \ket{\psi_4}=\frac{1}{\sqrt{2+\frac{k N_2}{l_1 N_1}}}\left[0,1,1,0,0,0,\sqrt{\frac{kN_2}{l_1N_1}}\right]^\intercal, \> \lambda_4=-1, \label{eq:onemarked_vecs} \\
	& \ket{\psi_5}=\frac{1}{\sqrt{1+\frac{l_2N_2}{l_1N_1}}}\left[0,0,0,1,0,0,\sqrt{\frac{l_2N_2}{l_1N_1}}\right]^\intercal, \> \lambda_5=-1, \nonumber \\
	& \ket{\psi_6}=\frac{1}{\sqrt{2+\frac{4(l_1N_1+l_2N_2)}{k N_2}}}\left[0,\frac{1}{\sqrt{2}},\frac{1}{\sqrt{2}},\sqrt{\frac{2l_2}{k}},-i\sqrt{\frac{2l_1N_1+(k+2l_2)N_2}{2kN_2}},i\sqrt{\frac{2l_1N_1+(k+2l_2)N_2}{2kN_2}},-\sqrt{\frac{2l_1 N_1}{k N_2}}\right]^\intercal, \> \lambda_6=-e^{i\phi}, \nonumber \\
	& \ket{\psi_7}=\frac{1}{\sqrt{2+\frac{4(l_1N_1+l_2N_2)}{k N_2}}}\left[0,\frac{1}{\sqrt{2}},\frac{1}{\sqrt{2}},\sqrt{\frac{2l_2}{k}},i\sqrt{\frac{2l_1N_1+(k+2l_2)N_2}{2kN_2}},-i\sqrt{\frac{2l_1N_1+(k+2l_2)N_2}{2kN_2}},-\sqrt{\frac{2l_1 N_1}{k N_2}}\right]^\intercal, \> \lambda_7=-e^{-i\phi}, \nonumber
\end{align}
\end{widetext}
where 
\begin{gather*}
	\sin \theta=\sqrt{\frac{2l_1 N_1+k N_2}{N_1N_2}}, \\
	\sin\phi=\sqrt{\frac{2l_1N_1+kN_2+2l_2N_2}{N_1N_2}}.
\end{gather*}
Now let us consider how the system evolves from each initial state, beginning with $\ket{s}$.


\subsection{Uniform Initial State Over Vertices}

\begin{figure}
\begin{center}
        \subfloat[] {
                \includegraphics{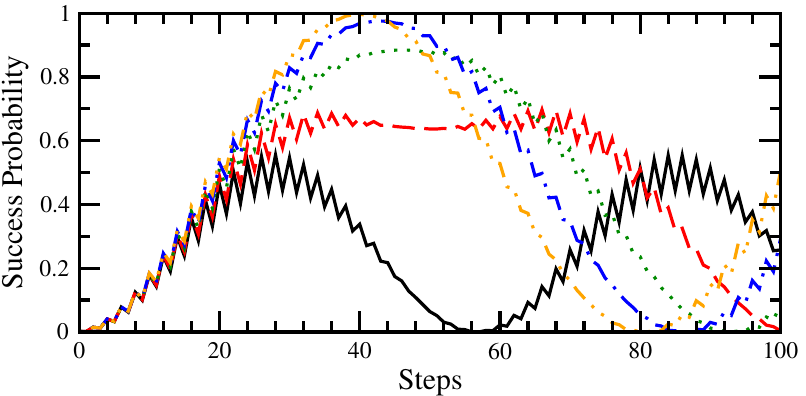}
                \label{fig:prob-s-1000-800-3-0a}
        }

	\subfloat[] {
                \includegraphics{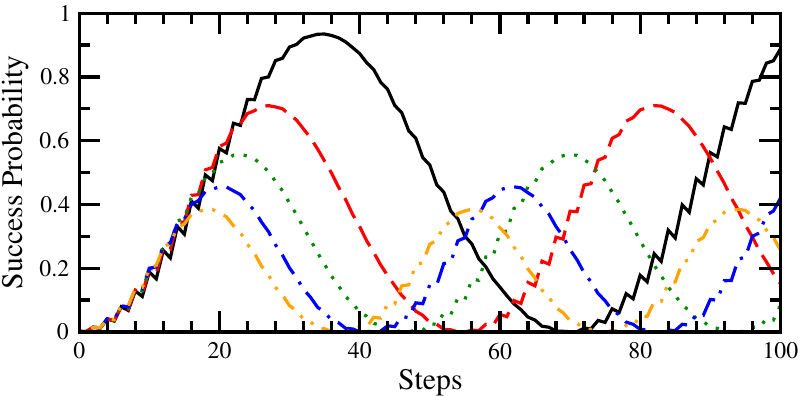}
                \label{fig:prob-s-1000-800-3-0b}
        }

	\subfloat[] {
                \includegraphics{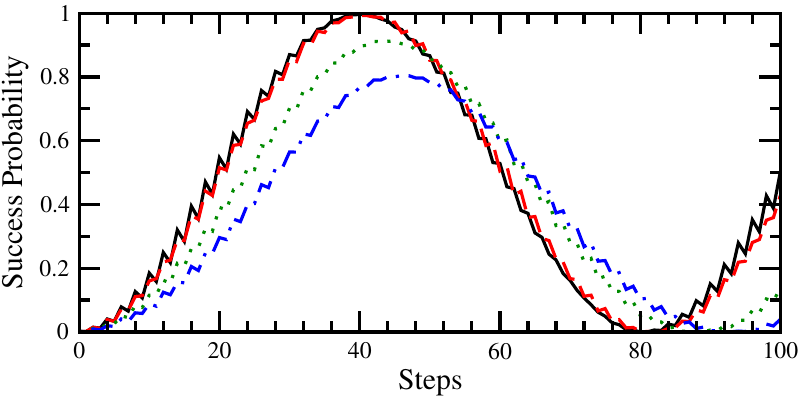}
                \label{fig:prob-s-1000-800-3-0c}
        }

	\subfloat[] {
		\includegraphics[width=3in]{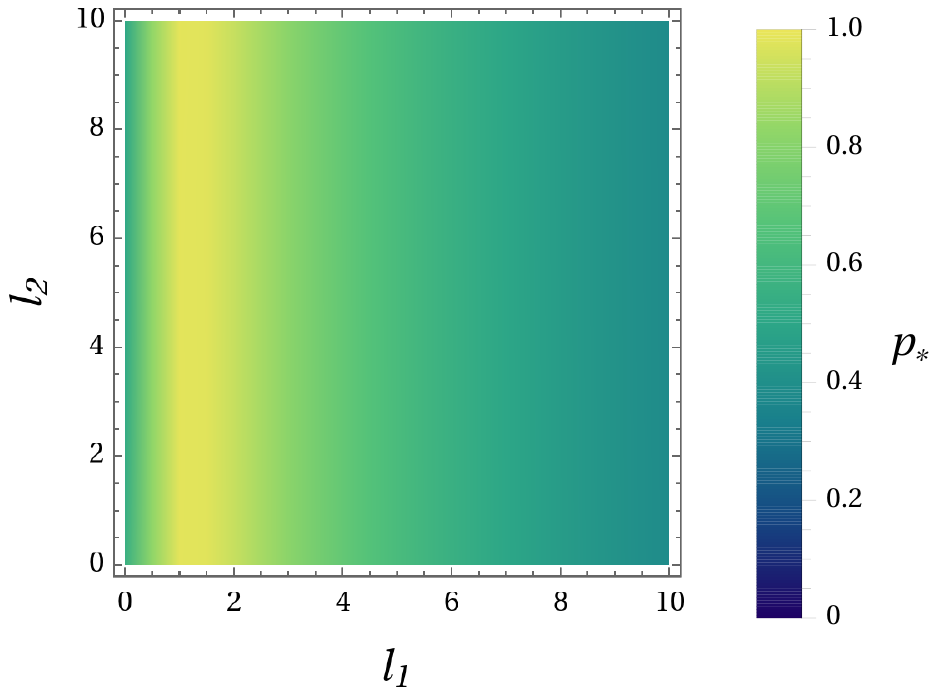}
		\label{fig:heatmap-s-1000-800-3-0}
        }

	\caption{\label{fig:prob-s-1000-800-3-0} Lackadaisical quantum search on the complete bipartite graph with $N_1=1000$, $N_2=800$, and $k=3$ marked vertices in set $X$, starting from initial state $\ket{s}$. In (a), $l_2 = 0$, and the solid black curve is $l_1 = 0$, dashed red is $0.05$, dotted green is $0.15$, and the dot-dashed blue is $0.25$. In (b), they are $l_2 = 0$ and $l_1 = 2, 4, 6, 8, 10$. In (c), they are $l_1 = 1.2$ and $l_2 = 0, 100, 1000, 2000$. (d) depicts the maximum success probability for various values of $l_1$ and $l_2$.} 
\end{center}
\end{figure}

In the 7D subspace, the uniform initial state over the vertices \eqref{eq:s} is 
\begin{align*}
	\ket{s}
		&= \frac{1}{\sqrt{N_1+N_2}} \bigg[ \sqrt{\frac{kl_1}{N_2+l_1}}\ket{aa} + \sqrt{\frac{kN_2}{N_2+l_1}}\ket{ab} \\
		&\quad+ \sqrt{\frac{kN_2}{N_1+l_2}}\ket{ba} + \sqrt{\frac{N_2l_2}{N_1+l_2}}\ket{bb} \\
		&\quad+ \sqrt{\frac{N_2(N_1-k)}{N_1+l_2}}\ket{bc} + \sqrt{\frac{N_2(N_1-k)}{N_2+l_1}}\ket{cb} \\
		&\quad+ \sqrt{\frac{l_1(N_1-k)}{N_2+l_1}}\ket{cc}\bigg].
\end{align*}
This evolves by repeated applications of $U$ \eqref{eq:onemarked_U}, and at time $t$, the success probability is
\[ p(t) = | \langle aa | U^t | s \rangle |^2 + | \langle ab | U^t | s \rangle |^2. \]
This is plotted in \fref{fig:prob-s-1000-800-3-0a} and \fref{fig:prob-s-1000-800-3-0b} with $N_1 = 1000$, $N_2 = 800$, $k = 3$, various values of $l_1$, and $l_2 = 0$. In \fref{fig:prob-s-1000-800-3-0a}, we see that as $l_1$ increases, the maximum success probability increases until it reaches a value near $1$ when $l_1 = 1.2$. In \fref{fig:prob-s-1000-800-3-0b}, the maximum success probability decreases as the weights are further increased. Thus, there is some optimal amount of ``laziness'' that maximizes the success probability, in this case $l_1 = 1.2$. In \fref{fig:prob-s-1000-800-3-0c}, we keep $l_1 = 1.2$ but now vary $l_2$. We see that as long as $l_2$ is small (compared to $N_1$ and $N_2$), it does not affect the evolution. This is also depicted in the heatmap in \fref{fig:heatmap-s-1000-800-3-0}, where we plot the maximum success probability for $l_1, l_2 \in [0,10]$. For these small values of $l_1$ and $l_2$, the maximum success probability only depends on $l_1$, not $l_2$.

\begin{figure}
\begin{center}
        \subfloat[] {
		\includegraphics{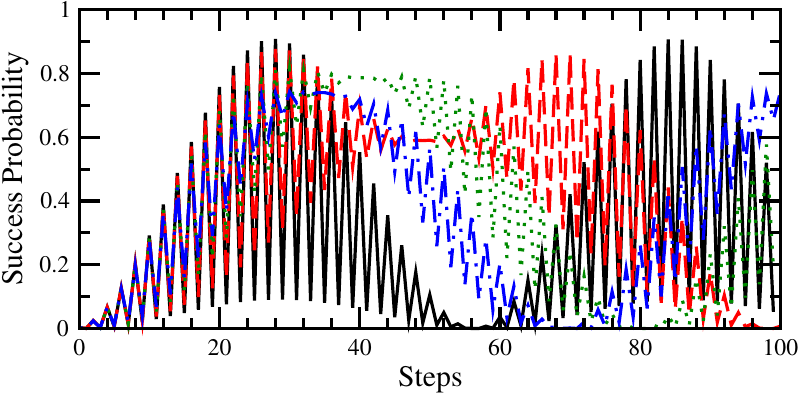}
		\label{fig:prob-s-1000-100-3-0} 
        }

	\subfloat[] {
		\includegraphics{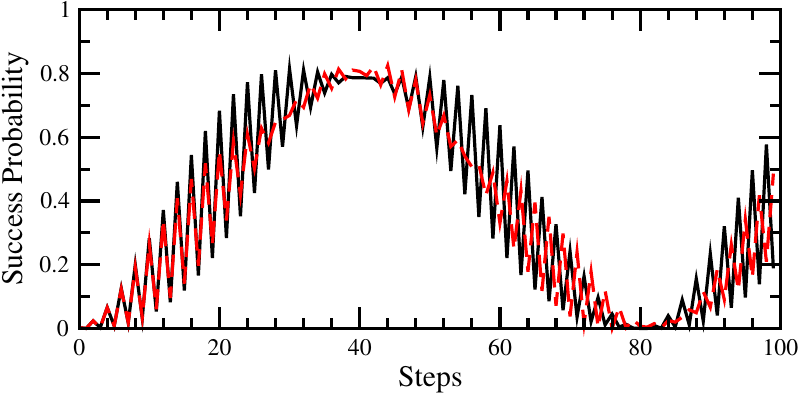}
		\label{fig:prob-s-1000-100-3-0-0_15} 
	}
	\caption{Lackadaisical quantum search on the complete bipartite graph with $N_1=1000$, $N_2=100$, and $k=3$ marked vertices in set $X$, starting from initial state $\ket{s}$. In (a), $l_2 = 0$, and the solid black curve is $l_1 = 0$, dashed red is $0.05$, dotted green is $0.15$, and dot-dashed blue is $0.25$. In (b), they are $l_1 = 0.15$ and $l_2 = 0, 3.137$.} 
\end{center}
\end{figure}

In \fref{fig:prob-s-1000-100-3-0}, we take $N_1 = 1000$ and $N_2 = 100$, so the graph is more irregular. From the solid black curve, we see that the loopless evolution oscillates wildly. This was proved in \cite{Wong31}. As $l_1$ is increased, the oscillations dampen, and a somewhat flat ``peak'' develops. This peak is greatest when $l_1 = 0.15$, corresponding to the dotted green curve. Its height, however, is worse than the loopless algorithm's strong oscillations. Thus, when the complete bipartite graph is highly irregular, the lackadaisical quantum walk yields no improvement. In \fref{fig:prob-s-1000-100-3-0-0_15}, we fix $l_1 = 0.15$ and plot $l_2 = 0$ and $3.137$. We see that increasing $l_2$ can also dampen the oscillations while slightly increasing the peak in success probability.

Next, we prove these results asymptotically. For large $N_1$ and $N_2$ (compared to $k$, $l_1$, and $l_2$), the $\ket{bc}$ and $\ket{cb}$ basis states dominate the initial state, so
\[ \ket{s}\approx\frac{1}{\sqrt{N_1+N_2}}\left(\sqrt{N_2}\ket{bc}+\sqrt{N_1}\ket{cb}\right). \]
We can express this as a linear combination of the asymptotic eigenvectors of $U$:
\[ \ket{s} =a \ket{\psi_1} + b \ket{\psi_2} + c \ket{\psi_3} + d \ket{\psi_4} + e \ket{\psi_5} + f \ket{\psi_6} + g \ket{\psi_7}, \]
where
\begin{align*}
	& a=-(\sqrt{N_1}+\sqrt{N_2})\sqrt{\frac{l_1N_1}{(N_1+N_2)(2l_1N_1+kN_2)}}, \\
	& b=c=\frac{\sqrt{k}N_2+\sqrt{kN_1N_2}}{2\sqrt{(N_1+N_2)(2l_1N_1+kN_2)}}, \\
	& d=e=0, \\
	& f=\frac{i(\sqrt{N_2}-\sqrt{N_1})}{2\sqrt{N_1+N_2}}, \\
	& g=\frac{i(\sqrt{N_1}-\sqrt{N_2})}{2\sqrt{N_1+N_2}}.
\end{align*}
Then each application of $U$ multiplies each eigenvector by its eigenvalue, so the system at time $t$ is
\begin{align*}
	U^t \ket{s}
		&= a \lambda_1^t \ket{\psi_1} + b \lambda_2^t \ket{\psi_2} + c \lambda_3^t \ket{\psi_3} \\
		&\quad+ f \lambda_6^t \ket{\psi_6} + g \lambda_7^t \ket{\psi_7}.
\end{align*}
The success probability at time $t$ is $p(t) = | \bra{aa} U^t \ket{s} |^2 + | \bra{ab} U^t \ket{s} |^2$. Evaluating this, we get
\begin{align*}
	p(t)
		&= \frac{1}{N_1+N_2} \Bigg\{ \bigg[ \frac{\sqrt{k N_2} (\sqrt{N_1} + \sqrt{N_2})}{2\sqrt{2l_1N_1+kN_2}} \sin(\theta t) \\
		&\quad + (-1)^t \frac{\sqrt{k N_2} (\sqrt{N_1} - \sqrt{N_2})}{2\sqrt{2l_1N_1+kN_2+2l_2N_2}} \sin(\phi t) \bigg]^2 \\
		&\quad+ \left[\frac{\sqrt{k l_1 N_1 N_2}(\sqrt{N_1} + \sqrt{N_2})}{2l_1N_1+kN_2} (\cos(\theta t)-1) \right]^2 \Bigg\}.
\end{align*}
This agrees well with \fref{fig:prob-s-1000-800-3-0} and \fref{fig:prob-s-1000-100-3-0}, since $N_1 = 1000$ and $N_2 = 800$ are large compared to the other parameters (except the dotted green and dot-dashed blue curves in \fref{fig:prob-s-1000-800-3-0c}).

Now let us prove what happens when $l_1$ takes the value
\begin{equation}
	\label{eq:onemarked_l1}
	l_1 = \frac{kN_2}{2N_1}.
\end{equation}
This corresponds to $l_1 = 1.2$ in \fref{fig:prob-s-1000-800-3-0} and $0.15$ in \fref{fig:prob-s-1000-100-3-0}.
Then $2l_1N_1 + kN_2 = 2kN_2$, and $p(t)$ simplifies to
\begin{align*}
	p(t)
		&= \frac{1}{N_1+N_2} \Bigg\{ \bigg[ \frac{\sqrt{N_1} + \sqrt{N_2}}{2\sqrt{2}} \sin(\theta t) \\
		&\quad + (-1)^t \frac{\sqrt{k} (\sqrt{N_1} - \sqrt{N_2})}{2\sqrt{2(k+l_2)}} \sin(\phi t) \bigg]^2 \\
		&\quad+ \left[\frac{\sqrt{N_1} + \sqrt{N_2}}{2\sqrt{2}} (\cos(\theta t)-1) \right]^2 \Bigg\}.
\end{align*}
At time
\begin{equation}
	\label{eq:onemarked_s_runtime}
	t_* = \frac{\pi}{\theta} \approx \frac{\pi}{\sin \theta} = \pi \sqrt{\frac{N_1N_2}{2l_1 N_1+k N_2}} = \pi \sqrt{\frac{N_1}{2k}},
\end{equation}
where we used \eqref{eq:onemarked_l1} in the last equality, and which does not depend on $N_2$, $l_1$, or $l_2$, the success probability reaches a value of $p_* = p(t_*)$, where
\begin{align*}
	p_*
		&= \frac{1}{N_1+N_2} \Bigg\{ \bigg[ \frac{\sqrt{k} (\sqrt{N_1} - \sqrt{N_2})}{2\sqrt{2(k+l_2)}} \sin \left( \frac{\pi\phi}{\theta} \right) \bigg]^2 \\
		&\quad+ \left[\frac{\sqrt{N_1} + \sqrt{N_2}}{\sqrt{2}} \right]^2 \Bigg\}.
\end{align*}
where we have taken $(-1)^{t_*} = \pm 1$ since the timesteps are integers. For large $N_1$ and $N_2$,
\begin{align*}
	\frac{\pi \phi}{\theta} 
		&\approx \frac{\pi \sin{\phi}}{\sin \theta} = \pi \sqrt{\frac{2l_1N_1 + k N_2 + 2l_2N_2}{2l_1N_1+kN_2}} = \pi \sqrt{1 + \frac{l_2}{k}}.
\end{align*}
Thus, at time $t_*$, the success probability reaches a value of
\begin{align}
	p_*
		&= \frac{1}{N_1+N_2} \Bigg\{ \bigg[ \frac{\sqrt{k} (\sqrt{N_1} - \sqrt{N_2})}{2\sqrt{2(k+l_2)}} \sin \left( \pi \sqrt{1 + \frac{l_2}{k}} \right) \bigg]^2 \nonumber \\
		&\quad+ \left[\frac{\sqrt{N_1} + \sqrt{N_2}}{\sqrt{2}} \right]^2 \Bigg\}. \label{eq:onemarked_s_prob}
\end{align}
Plugging in the parameters for the dot-dot-dashed orange curve from \fref{fig:prob-s-1000-800-3-0a}, our equations indicate when $l_1 = kN_2 / 2N_1 = 1.2$ \eqref{eq:onemarked_l1} and $l_2 = 0$, then at time $t_* = 41$ \eqref{eq:onemarked_s_runtime}, the success probability is $p_* = 0.996904$ \eqref{eq:onemarked_s_prob_lower}, in great agreement with the figure. When $l_2 = 10$, we instead get $p_* = 0.996915$, so the difference in success probability is minimal, as shown in \fref{fig:heatmap-s-1000-800-3-0}. Next, using the numbers from \fref{fig:prob-s-1000-100-3-0-0_15}, when $l_1 = 0.15$ and $l_2 = 0$, then at time $t_* = 41$, the success probability is $p_* = 0.78748$, whereas when $l_2 = 3.13725$, $p_* = 0.812225$, which is a noticeable increase in success probability.

To find the value of $l_2$ that maximizes $p_*$, we take the derivative of \eqref{eq:onemarked_s_prob} and set it equal to zero, which upon simplifying, yields the following equation for $l_2$:
\[ \pi \sqrt{1 + \frac{l_2}{k}} = \tan \left( \pi \sqrt{1 + \frac{l_2}{k}} \right). \]
Numerically solving this transcendental equation yields
\[ \sqrt{1 + \frac{l_2}{k}} = 1.4303, \]
or
\[ l_2 = 1.0458\,k. \]
With this value of $l_2$, the maximum success probability \eqref{eq:onemarked_s_prob} is
\begin{equation}
	\label{eq:onemarked_s_prob_upper}
	p_* = \frac{0.1164 \left( \sqrt{N_1} - \sqrt{N_2} \right)^2 + \left( \sqrt{N_1} + \sqrt{N_2} \right)^2}{2(N_1 + N_2)}.
\end{equation}
With $k = 3$, the optimal value of $l_2$ is $3.13725$, which is why it was used in \fref{fig:prob-s-1000-100-3-0-0_15}. As previously stated, its maximum success probability is 0.812225, which is less than the loopless algorithm's $\max(N_1,N_2)/(N_1 + N_2) = 1000/(1000+100) = 0.90909$ \cite{Wong31}. Thus, the lackadaisical quantum walk performs worse with these parameters.

Let us explore when the lackadaisical quantum walk does yield a better success probability than the loopless algorithm. Symbolically, it is difficult to compare \eqref{eq:onemarked_s_prob} and \eqref{eq:onemarked_s_prob_upper} with $\max(N_1,N_2)/(N_1 + N_2)$. To make the comparison more direct, we can lower-bound $p_*$ \eqref{eq:onemarked_s_prob} by taking $\sin(\cdot)$ to be zero, which occurs exactly when $l_2 = 0$. This yields
\begin{equation}
	p_* \ge \frac{1}{N_1+N_2} \left[\frac{\sqrt{N_1} + \sqrt{N_2}}{\sqrt{2}} \right]^2 = \frac{(\sqrt{N_1} + \sqrt{N_2})^2}{2(N_1 + N_2)} \label{eq:onemarked_s_prob_lower}.
\end{equation}
Then the lackadaisical quantum walk reaches a higher success probability than the loopless algorithm when
\[ \frac{(\sqrt{N_1} + \sqrt{N_2})^2}{2(N_1 + N_2)} > \frac{\max(N_1,N_2)}{N_1 + N_2}. \]
Assuming $N_1 > N_2$, this occurs when
\[ N_2 > (3 - 2\sqrt{2}) N_1 \approx 0.172 \, N_1. \]
Using $N_1 = 1000$, this proves that the success probability is improved when $N_2 > 172$, regardless of the value of $l_2$, which is satisfied in \fref{fig:prob-s-1000-800-3-0} with $N_2 = 800$, but is not satisfied in \fref{fig:prob-s-1000-100-3-0} with $N_2 = 100$.

If the complete bipartite graph is regular, then $N_1 = N_2 = N/2$, and so $l_1 = k/2$ \eqref{eq:onemarked_l1}, $t_* = (\pi/2) \sqrt{N/k}$ \eqref{eq:onemarked_s_runtime}, and $p_* = 1$ \eqref{eq:onemarked_s_prob}, for any value of $l_2$. By comparison, from \cite{Wong31}, the non-lackadaisical (loopless) algorithm reaches a success probability of $1/2$ at time $(\pi/2\sqrt{2}) \sqrt{N/k}$, and since we must repeat the algorithm twice, on average, before finding a marked vertex, the total runtime is $(\pi/\sqrt{2}) \sqrt{N/k}$. Thus, the lackadaisical quantum walk yields a constant-factor improvement of $1/\sqrt{2}$ over the loopless algorithm, the same improvement as search on the complete graph \cite{Wong10}.


\subsection{Stationary Initial State Under Walk}

\begin{figure}
\begin{center}
        \subfloat[] {
                \includegraphics{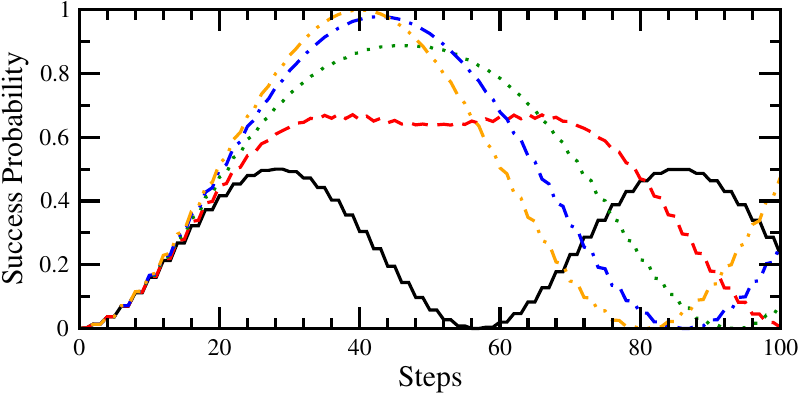}
                \label{fig:prob-sigma-1000-800-3-0a}
        }

	\subfloat[] {
                \includegraphics{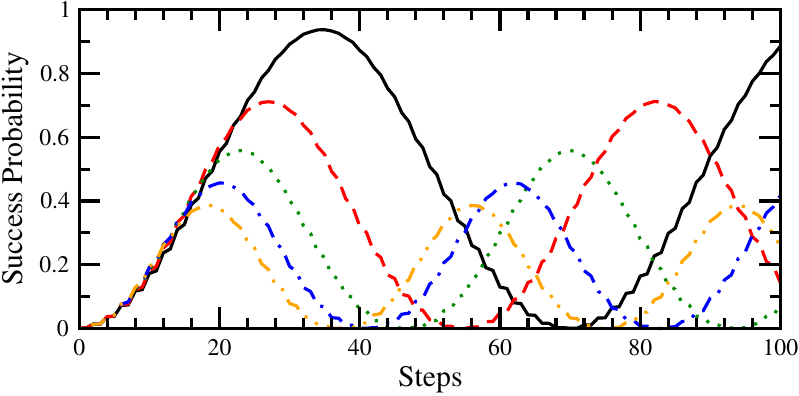}
                \label{fig:prob-sigma-1000-800-3-0b}
        }

	\subfloat[] {
                \includegraphics{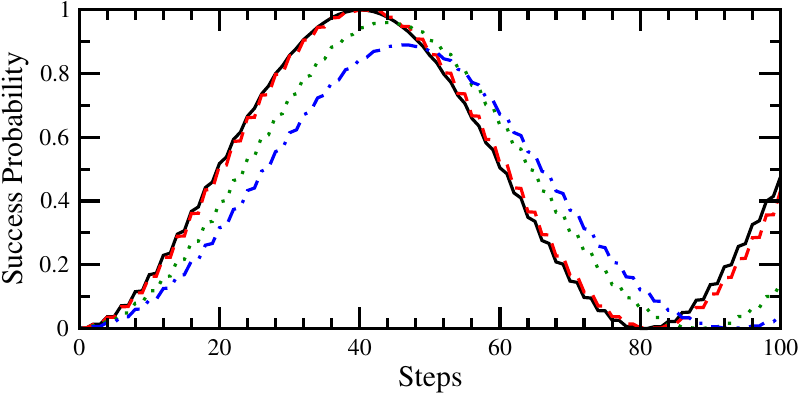}
                \label{fig:prob-sigma-1000-800-3-0c}
        }

	\subfloat[] {
		\includegraphics[width=3in]{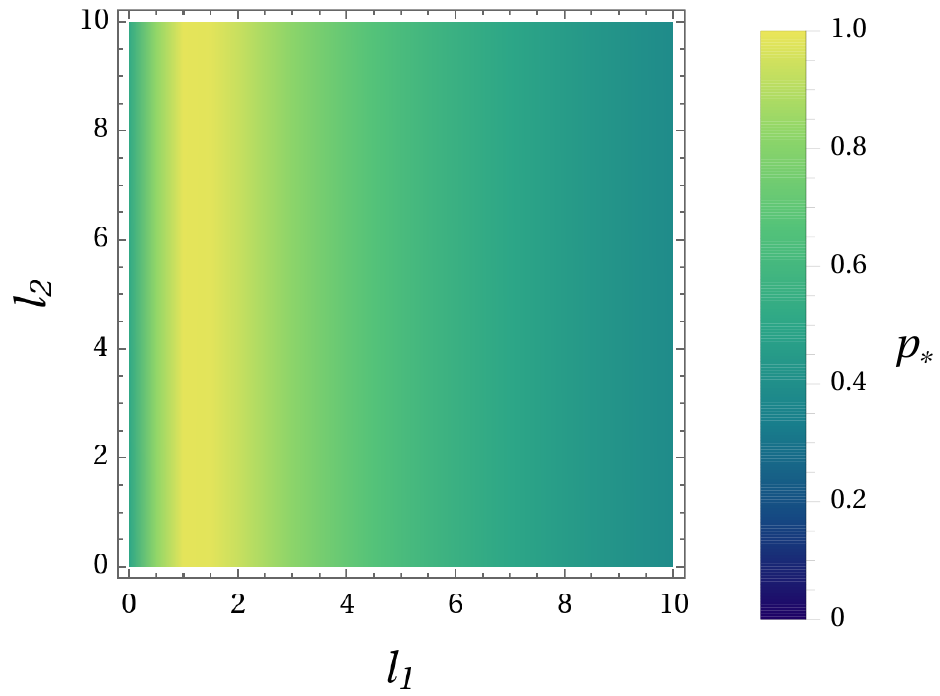}
		\label{fig:heatmap-sigma-1000-800-3-0}
        }

	\caption{\label{fig:prob-sigma-1000-800-3-0} Lackadaisical quantum search on the complete bipartite graph with $N_1=1000$, $N_2=800$, and $k=3$ marked vertices in set $X$, starting from initial state $\ket{\sigma}$. In (a), $l_2 = 0$, and the solid black curve is $l_1 = 0$, dashed red is $0.3$, dotted green is $0.6$, dot-dashed blue is $0.9$, and dot-dot-dashed orange is $1.2$. In (b), they are $l_2 = 0$ and $l_1 = 2, 4, 6, 8, 10$. In (c), they are $l_1 = 1.2$ and $l_2 = 0, 100, 1000, 2000$. (d) depicts the maximum success probability for various values of $l_1$ and $l_2$.} 
\end{center}
\end{figure}

\begin{figure}
\begin{center}
	\includegraphics{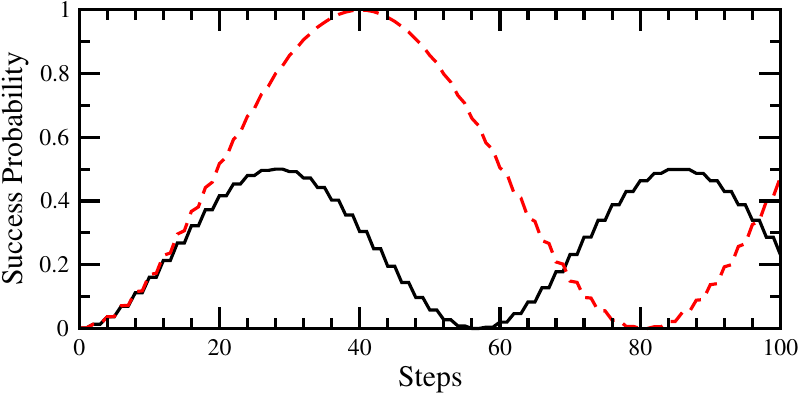}
	\caption{\label{fig:prob-sigma-1000-100-3-0} Success probability as a function of time for lackadaisical search on the complete bipartite graph with $N_1=1000$, $N_2=100$, and $k=3$ marked vertices in set $X$, starting from initial state $\ket{\sigma}$. The solid black curve is $l_1 = l_2 = 0$, and the dashed red is $l_1 = 0.15$ and $l_2 = 0$.}
\end{center}
\end{figure}

Now let us consider search when the initial state is stationary under the quantum walk. In the 7D subspace, the initial state \eqref{eq:sigma} is
\begin{align*}
	\ket{\sigma}
		&= \frac{1}{\sqrt{2N_1N_2+l_1N_1+l_2N_2}} \big[ \sqrt{kl_1} \ket{aa} + \sqrt{kN_2} \ket{ab} \\
		&\quad+ \sqrt{kN_2} \ket{ba} + \sqrt{l_2N_2} \ket{bb} + \sqrt{N_2(N_1-k)} \ket{bc} \\
		&\quad+ \sqrt{N_2(N_1-k)} \ket{cb} + \sqrt{l_1(N_1-k)} \ket{cc} \big].
\end{align*}
Then repeatedly multiplying by $U$ \eqref{eq:onemarked_U}, the success probability at time $t$ is $p(t) = | \bra{aa} U^t \ket{\sigma} |^2 + | \bra{ab} U^t \ket{\sigma} |^2$. This is plotted in \fref{fig:prob-sigma-1000-800-3-0} with the same parameters as \fref{fig:prob-s-1000-800-3-0}, but with $\ket{\sigma}$ instead of $\ket{s}$. They evolve very similarly, except $\ket{\sigma}$'s evolution is smoother. Again, the maximum success probability increases until $l_1 = 1.2$, and then it decreases, and $l_2$ has negligible effect on the evolution as long as it is small. In \fref{fig:prob-sigma-1000-100-3-0}, we plot $p(t)$ with the same parameters as \fref{fig:prob-s-1000-100-3-0}. We see that even though the graph is highly irregular, the success probability is still boosted from $1/2$ to $1$. It was proven in \cite{Wong31} that the loopless evolution always reaches $1/2$. Now let us prove that the lackadaisical quantum walk boosts it to $1$.

For large values of $N_1$ and $N_2$, the initial state is asymptotically
\[ \ket{\sigma} = \frac{1}{\sqrt{2}}(\ket{bc}+\ket{cb}). \]
Expressing this as a linear combination of the eigenvectors \eqref{eq:onemarked_vecs} of $U$,
\[ \ket{\sigma} = a \ket{\psi_1} + b \ket{\psi_2} + c \ket{\psi_3} + d \ket{\psi_4} + e \ket{\psi_5} + f \ket{\psi_6} + g \ket{\psi_7}, \]
where
\begin{align*}
	&a = -\sqrt{\frac{2 l_1 N_1}{2 l_1 N_1 + k N_2}}, \\
	&b = c = \sqrt{\frac{k N_2}{2(2 l_1 N_1 + k N_2)}}, \\
	&d = e = f = g = 0.
\end{align*}
Then the system at time $t$ is
\[ U^t \ket{\sigma} = a\lambda_1^t\ket{\psi_1} + b\lambda_2^t\ket{\psi_2} + c\lambda_3^t\ket{\psi_3}. \]
Taking the inner product with $\ket{aa}$ and $\ket{ab}$, squaring, and summing, the success probability at time $t$ is
\begin{align}
	p(t) 
		&= \frac{k N_2 \left[ 6 l_1 N_1 + k N_2 + (k N_2 - 2 l_1 N_1) \cos(\theta t) \right]}{(2 l_1 N_1 + k N_2)^2} \nonumber \\
		&\quad\times \sin^2 \left( \frac{\theta t}{2} \right). \label{eq:onemarked_sigma_p}
\end{align}
Now when $l_1 = kN_2/2N_1$ \eqref{eq:onemarked_l1}, this becomes
\[ p(t) = \sin^2 \left( \frac{\theta t}{2} \right), \]
and so the success probability reaches
\begin{equation}
	\label{eq:onemarked_sigma_prob_1}
	p_* = 1
\end{equation}
at time $t_* = \pi/\theta \approx \pi \sqrt{N_1/2k}$ \eqref{eq:onemarked_s_runtime}. So using the initial state $\ket{\sigma}$, the lackadaisical quantum walk always boosts the success probability from $1/2$ to $1$. 

Although $l_1 = kN_2/2N_1$ \eqref{eq:onemarked_l1} maximizes the peak success probability to 1, we can also find the peak success probability for any value of $l_1$ (small compared to $N_1$ and $N_2$). Taking the derivative of $p(t)$ \eqref{eq:onemarked_sigma_p},
\[ \frac{dp}{dt} = \frac{k N_2 \theta \sin (\theta t) [ \cos (\theta  t) (k N_2 - 2 l_1 N_1) + 4 l_1 N_2]}{(k N_2 + 2 l_1 N_1)^2}. \]
Setting this equal to zero, the runtime is
\begin{equation}
	\label{eq:onemarked_sigma_runtime}
	t_* = \begin{cases}
		\frac{1}{\theta} \cos^{-1} \left( \frac{4 l_1 N_1}{2 l_1 N_2 - k N_2} \right) & 0 < l_1 \leq \frac{k N_2}{6 N_1}, \\
		\frac{\pi}{\theta} & l_1 > \frac{k N_2}{6 N_1}. \\
	\end{cases}
\end{equation}
At this runtime, the maximum success probability is $p_* = p(t_*)$:
\begin{equation}
	\label{eq:onemarked_sigma_prob}
	p_* = \begin{cases}
		\frac{kN_2}{2kN_2 - 4l_1 N_1} & 0 < l_1 \leq \frac{k N_2}{6 N_1}, \\
		\frac{8 k l_1 N_1 N_2}{(2 l_1 N_1 + k N_2)^2}, & l_1 > \frac{k N_2}{6 N_1}. \\
	\end{cases}
\end{equation}
These expressions for $t_*$ and $p_*$ are consistent with Figs.~\ref{fig:prob-sigma-1000-800-3-0a} and \ref{fig:prob-sigma-1000-800-3-0b}. Note all these results are independent of $l_2$, so long as $N_1$ and $N_2$ are large by comparison.

Finally, consider the case when the complete bipartite graph is regular, so $N_1=N_2=N/2$. Then $\ket{s} = \ket{\sigma}$. Recall without self-loops, the quantum walk achieves a success probability of $1/2$ \cite{Wong31}. The lackadaisical quantum walk is better than this when $p_* > 1/2$. Solving this inequality along with \eqref{eq:onemarked_sigma_prob}, we get
\[ 0 < l_1 < \frac{(3+2\sqrt{2})k}{2}. \]
So for this range of values of $l_1$, the lackadaisical quantum walk has a success probability better than $1/2$.

For comparison, non-lackadaisical search on the complete graph behaves the same way as the complete bipartite graph when $k = 1$ and $N_1 = N_2 = N/2$, with both graphs supporting a success probability of $1/2$ at time $(\pi/2\sqrt{2})\sqrt{N}$. As shown in \cite{Wong27}, lackadaisicial search on the complete graph has a better success probability than $1/2$ when $0 < l < 3+2\sqrt{2}$, where each vertex has a self-loop of weight $l$. Here, we have shown that on the complete bipartite graph, it has a better success probability when $0 < l_1 < (3+2\sqrt{2})/2$, which is half the range of weights.


\section{Marked Vertices in Both Sets}

\begin{figure}
\begin{center}
	\includegraphics{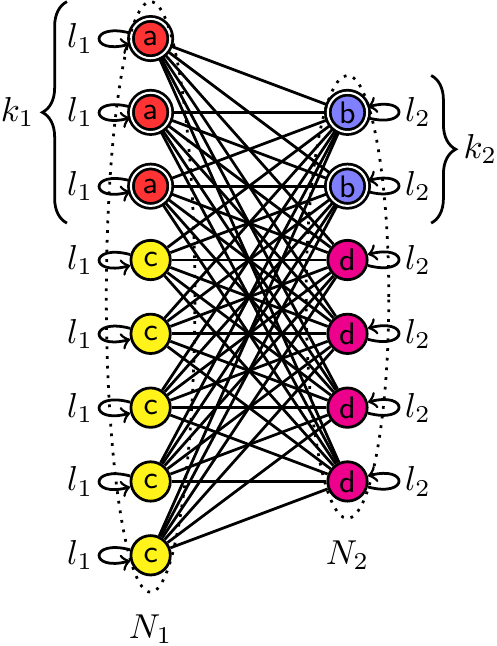}
	\caption{\label{fig:bipartite_bothmarked}The complete bipartite graph containing $N_1$ vertices in set $X$, each with a self-loop of weight $l_1$, and $N_2$ vertices in set $Y$, each with a self-loop of weight $l_2$. There are $k_1$ marked vertices in set $X$ and $k_2$ marked vertices in set $Y$. Marked vertices are indicated by double circles. Vertices that evolve identically share the same color and letter, and the letters correspond to the states of the subspace basis vectors.}
\end{center}
\end{figure}

Now we consider the case when there are marked vertices in both partite sets. As shown in \fref{fig:bipartite_bothmarked}, say there are $k_1$ marked vertices in set $X$ and $k_2$ marked vertices in set $Y$. Then regardless of whether the system starts in $\ket{s}$ or $\ket{\sigma}$, it evolves in a 12D subspace spanned by the following orthonormal basis states:
\begin{align*}
	& \ket{aa}=\frac{1}{\sqrt{k_1}}\sum_a \ket{a}\otimes\ket{a}, \\
	& \ket{ab}=\frac{1}{\sqrt{k_1}}\sum_a \ket{a}\otimes\frac{1}{\sqrt{k_2}}\sum_b \ket{b}, \\
	& \ket{ad}=\frac{1}{\sqrt{k_1}}\sum_a \ket{a}\otimes\frac{1}{\sqrt{N_2-k_2}}\sum_d \ket{d}, \\ 
	& \ket{ba}=\frac{1}{\sqrt{k_2}}\sum_b \ket{b}\otimes\frac{1}{\sqrt{k_1}}\sum_a \ket{a}, \\
	& \ket{bb}=\frac{1}{\sqrt{k_2}}\sum_b \ket{b}\otimes\ket{b}, \\
	& \ket{bc}=\frac{1}{\sqrt{k_2}}\sum_b \ket{b}\otimes\frac{1}{\sqrt{N_1-k_1}}\sum_c \ket{c}, \\
	& \ket{cb}=\frac{1}{\sqrt{N_1-k_1}}\sum_c \ket{c}\otimes\frac{1}{\sqrt{k_2}}\sum_b \ket{b}, \\
	& \ket{cc}=\frac{1}{\sqrt{N_1-k_1}}\sum_c \ket{c}\otimes\ket{c}, \\
	& \ket{cd}=\frac{1}{\sqrt{N_1-k_1}}\sum_c \ket{c}\otimes\frac{1}{\sqrt{N_2-k_2}}\sum_d \ket{d}, \\
	& \ket{da}=\frac{1}{\sqrt{N_2-k_2}}\sum_d \ket{d}\otimes\frac{1}{\sqrt{k_1}}\sum_a \ket{a}, \\
	& \ket{dc}=\frac{1}{\sqrt{N_2-k_2}}\sum_d \ket{d}\otimes\frac{1}{\sqrt{N_1-k_1}}\sum_c \ket{c}, \\
	& \ket{dd}=\frac{1}{\sqrt{N_2-k_2}}\sum_d \ket{d}\otimes\ket{d}.
\end{align*}

In this 12D subspace, initial states $\ket{s}$ \eqref{eq:s} and $\ket{\sigma}$ \eqref{eq:sigma} are
\begin{align*}
	\ket{s}
		&= \frac{1}{\sqrt{N_1+N_2}} \bigg[ \sqrt{\frac{k_1l_1}{N_2+l_1}}\ket{aa}+\sqrt{\frac{k_1k_2}{N_2+l_1}}\ket{ab} \\
		&\quad+ \sqrt{\frac{k_1(N_2-k_2)}{N_2+l_1}}\ket{ad} + \sqrt{\frac{k_1k_2}{N_1+l_2}}\ket{ba} \\
		&\quad+ \sqrt{\frac{k_2l_2}{N_1+l_2}}\ket{bb} + \sqrt{\frac{k_2(N_1-k_1)}{N_1+l_2}}\ket{bc} \\
		&\quad+ \sqrt{\frac{k_2(N_1-k_1)}{N_2+l_1}}\ket{cb} + \sqrt{\frac{l_1(N_1-k_1)}{N_2+l_1}}\ket{cc} \\
		&\quad+ \sqrt{\frac{(N_1-k_1)(N_2-k_2)}{N_2+l_1}}\ket{cd} + \sqrt{\frac{k_1(N_2-k_2)}{N_1+l_2}}\ket{da} \\
		&\quad+ \sqrt{\frac{(N_1-k_1)(N_2-k_2)}{N_1+l_2}}\ket{dc} + \sqrt{\frac{l_2(N_2-k_2)}{N_1+l_2}}\ket{dd}\bigg]
\end{align*}
and
\begin{align*}
	\ket{\sigma} &=
		\frac{1}{\sqrt{2N_1N_2+l_1N_1+l_2N_2}} \bigg[ \sqrt{k_1l_1}\ket{aa} + \sqrt{k_1k_2}\ket{ab} \\
		&\quad+ \sqrt{k_1(N_2-k_2)}\ket{ad} + \sqrt{k_1k_2}\ket{ba} + \sqrt{k_2l_2}\ket{bb} \\
		&\quad+ \sqrt{k_2(N_1-k_1)}\ket{bc} + \sqrt{k_2(N_1-k_1)}\ket{cb} \\
		&\quad+ \sqrt{l_1(N_1-k_1)}\ket{cc} + \sqrt{(N_1-k_1)(N_2-k_2)}\ket{cd} \\
		&\quad+ \sqrt{k_1(N_2-k_2)}\ket{da} + \sqrt{(N_1-k_1)(N_2-k_2)}\ket{dc} \\
		&\quad+ \sqrt{l_2(N_2-k_2)}\ket{dd} \bigg].
\end{align*}
The system evolves through repeated applications of $U=SCQ$, which in the 12D basis is
\begin{equation}
	\label{eq:U_bothmarked}
	U = \begin{pmatrix}
		U_1 & U_2 & U_3 \\
		U_4 & U_5 & U_6 \\
		U_7 & U_8 & U_9 \\
	\end{pmatrix},
\end{equation}
where
\begin{gather*}
	U_1 = \begin{pmatrix}
		\frac{N_2-l_1}{N_2+l_1} & \frac{-2\sqrt{k_2l_1}}{N_2+l_1} & \frac{-2\sqrt{l_1(N_2-k_2)}}{N_2+l_1} & 0 \\ 
		0 & 0 & 0 & \frac{N_1+l_2-2k_1}{N_1+l_2} \\
		0 & 0 & 0 & 0 \\
		\frac{-2\sqrt{k_2l_1}}{N_2+l_1} & \frac{N_2+l_1-2k_2}{N_2+l_1} & \frac{-2\sqrt{k_2(N_2-k_2)}}{N_2+l_1} & 0 \\ 
	\end{pmatrix}, \\
	U_2 = \begin{pmatrix}
		0 & 0 & 0 & 0 \\
		\frac{-2\sqrt{k_1l_2}}{N_1+l_2} & \frac{-2\sqrt{k_1(N_1-k_1)}}{N_1+l_2} & 0 & 0 \\ 
		0 & 0 & 0 & 0 \\
		0 & 0 & 0 & 0 \\ 
	\end{pmatrix}, \\
	U_3 = \begin{pmatrix}
		0 & 0 & 0 & 0 \\
		0 & 0 & 0 & 0 \\
		0 & \frac{2k_1-N_1-l_2}{N_1+l_2} & \frac{2\sqrt{k_1(N_1-k_1)}}{N_1+l_2} & \frac{2\sqrt{k_1l_2}}{N_1+l_2} \\
		0 & 0 & 0 & 0 \\
	\end{pmatrix}, \\
	U_4 = \begin{pmatrix}
		0 & 0 & 0 & \frac{-2\sqrt{k_1l_2}}{N_1+l_2} \\
		0 & 0 & 0 & 0 \\
		0 & 0 & 0 & \frac{-2\sqrt{k_1(N_1-k_1)}}{N_1+l_2} \\
		0 & 0 & 0 & 0 \\
	\end{pmatrix}, \\
	U_5 = \begin{pmatrix}
		\frac{N_1-l_2}{N_1+l_2} & \frac{-2\sqrt{l_2(N_1-k_1)}}{N_1+l_2} & 0 & 0 \\
		0 & 0 & \frac{2k_2-N_2-l_1}{N_2+l_1} & \frac{2\sqrt{k_2l_1}}{N_2+l_1} \\
		\frac{-2\sqrt{l_2(N_1-k_1)}}{N_1+l_2} & \frac{2k_1+l_2-N_1}{N_1+l_2}& 0 & 0 \\
		0 & 0 & \frac{2\sqrt{k_2l_1}}{N_2+l_1} & \frac{l_1-N_2}{N_2+l_1} 
	\end{pmatrix}, \\
	U_6 = \begin{pmatrix}
		0 & 0 & 0 & 0 \\
		\frac{2\sqrt{k_2(N_2-k_2)}}{N_2+l_1} & 0 & 0 & 0 \\
		0 & 0 & 0 & 0 \\
		\frac{2\sqrt{l_1(N_2-k_2)}}{N_2+l_1} & 0 & 0 & 0 \\
	\end{pmatrix}, \\
	U_7 = \begin{pmatrix}
		0 & 0 & 0 & 0 \\
		\frac{-2\sqrt{l_1(N_2-k_2)}}{N_2+l_1} & \frac{-2\sqrt{k_2(N_2-k_2)}}{N_2+l_1} & \frac{2k_2+l_1-N_2}{N_2+l_1} & 0 \\
		0 & 0 & 0 & 0 \\ 
		0 & 0 & 0 & 0 \\ 
	\end{pmatrix}, \\
	U_8 = \begin{pmatrix}
		0 & 0 & 0 & 0 \\ 
		0 & 0 & 0 & 0 \\ 
		0 & 0 & \frac{2\sqrt{k_2(N_2-k_2)}}{N_2+l_1} & \frac{2\sqrt{l_1(N_2-k_2)}}{N_2+l_1} \\
		0 & 0 & 0 & 0 \\ 
	\end{pmatrix}, \\
	U_9 = \begin{pmatrix}
		0 & \frac{2\sqrt{k_1(N_1-k_1)}}{N_1+l_2} & \frac{N_1-2k_1-l_2}{N_1+l_2} & \frac{2\sqrt{l_2(N_1-k_1)}}{N_1+l_2} \\
		0 & 0 & 0 & 0 \\
		\frac{N_2-2k_2-l_1}{N_2+l_1} & 0 & 0 & 0 \\
		0 & \frac{2\sqrt{k_1l_2}}{N_1+l_2} & \frac{2\sqrt{l_2(N_1-k_1)}}{N_1+l_2} & \frac{l_2-N_1}{N_1+l_2}
	\end{pmatrix}.
\end{gather*}

In Appendix~\ref{sec:eigen_bothmarked}, we attempted to use perturbation theory to asymptotically determine the eigenvectors of $U$ for large $N_1$ and $N_2$. This was too complicated in general, however. As a result, most of our investigation will be numerical. Yet, for the special case when the graph is regular ($N_1 = N_2 = N/2$), suggesting that the weights of the self-loops should be identical ($l_1 = l_2 = l$), and assuming the number of marked vertices in both sets is equal ($k_1 = k_2 = k/2$), half of the eigenvectors could be found. They are
\begin{widetext}
\begin{equation}
	\label{eq:vecs_bothmarked}
	\begin{aligned}
		& \ket{\psi_1}=\frac{1}{\sqrt{4+\frac{4k}{l}}}\left[1,\sqrt{\frac{k}{2l}},\frac{i\sqrt{k+l}}{\sqrt{2l}},\sqrt{\frac{k}{2l}},1,\frac{i\sqrt{k+l}}{\sqrt{2l}},\frac{-i\sqrt{k+l}}{\sqrt{2l}},0,\sqrt{\frac{k}{2l}},\frac{-i\sqrt{k+l}}{\sqrt{2l}},\sqrt{\frac{k}{2l}},0\right]^\intercal, \> \lambda_1=e^{-i\theta}, \\
		& \ket{\psi_2}=\frac{1}{\sqrt{4+\frac{4k}{l}}}\left[1,\sqrt{\frac{k}{2l}},\frac{-i\sqrt{k+l}}{\sqrt{2l}},\sqrt{\frac{k}{2l}},1,\frac{-i\sqrt{k+l}}{\sqrt{2l}},\frac{i\sqrt{k+l}}{\sqrt{2l}},0,\sqrt{\frac{k}{2l}},\frac{i\sqrt{k+l}}{\sqrt{2l}},\sqrt{\frac{k}{2l}},0\right]^\intercal, \> \lambda_2=e^{i\theta}, \\
		& \ket{\psi_3}=\frac{1}{2}\left[1,0,\frac{i}{\sqrt{2}},0,-1,\frac{-i}{\sqrt{2}},\frac{i}{\sqrt{2}},0,0,\frac{-i}{\sqrt{2}},0,0\right]^\intercal, \> \lambda_3=e^{-i\phi}, \\
		& \ket{\psi_4}=\frac{1}{2}\left[1,0,\frac{-i}{\sqrt{2}},0,-1,\frac{i}{\sqrt{2}},\frac{-i}{\sqrt{2}},0,0,\frac{i}{\sqrt{2}},0,0\right]^\intercal, \> \lambda_4=e^{i\phi}, \\
		& \ket{\psi_5}=\frac{1}{\sqrt{2+\frac{4l}{k}}}\left[1,0,0,0,1,0,0,0,-\sqrt{\frac{2l}{k}},0,-\sqrt{\frac{2l}{k}},0\right]^\intercal, \> \lambda_5=1, \\
		& \ket{\psi_6}=\frac{1}{\sqrt{2}}\left[0,\frac{1}{\sqrt{2}},0,\frac{1}{\sqrt{2}},0,0,0,0,\frac{-1}{\sqrt{2}},0,\frac{-1}{\sqrt{2}},0\right]^\intercal, \> \lambda_6=1,
	\end{aligned}
\end{equation}
\end{widetext}
where
\[ \sin\theta=\frac{2\sqrt{k+l}}{\sqrt{N}}, \quad\text{and}\quad\sin\phi=\frac{2\sqrt{l}}{\sqrt{N}}. \]
The remaining six eigenvectors, which to leading-order are degenerate with eigenvalue $-1$, are more complicated, but they are not necessary since the initial states can be expressed as linear combinations of $\ket{\psi_1}, \dots, \ket{\psi_6}$ given above. Next, let us focus on this specific case, and afterward, we will numerically explore the search problem more generally.


\subsection{Symmetric Search Problem}

\begin{figure}
\begin{center}
	\includegraphics{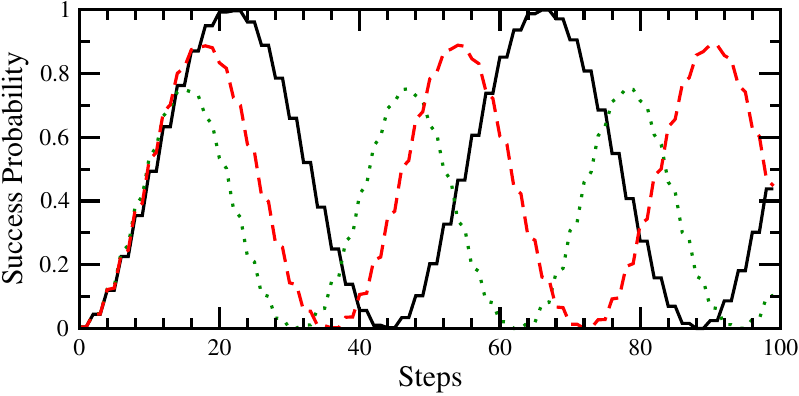}
	\caption{\label{fig:prob-1000-1000-5-5-symmetric} Lackadaisical quantum search on the complete bipartite graph with $N_1 = N_2 = 1000$, $l_1 = l_2 = l$, and $k_1 = k_2 = 5$ from either initial state $\ket{s} = \ket{\sigma}$. The solid black curve is $l = 0$, the dashed red curve is $l = 5$, and the dotted green curve is $l = 10$.}
\end{center}
\end{figure}

Say the search problem on the complete bipartite graph is symmetric, meaning the two partite sets have the same structure. Thus, the graph is regular ($N_1 = N_2 = N/2$), and the self-loops all have the same weight ($l_1 = l_2 = l$). Then the initial states are equal: $\ket{s} = \ket{\sigma} = \ket{\psi(0)}$. For the sets to have the same structure, we also assume each contains the same number of marked vertices ($k_1 = k_2 = k/2$). Then the initial state is
\begin{align*}
	\ket{\psi(0)} 
		&= \frac{1}{\sqrt{2N(N+2l)}} \Big[ \sqrt{2kl} \ket{aa} + k \ket{ab} \\
		&\quad+ \sqrt{k(N-k)} \ket{ad} + k \ket{ba} + \sqrt{2kl} \ket{bb} \\
		&\quad+ \sqrt{k(N-k)} \ket{bc} + \sqrt{k(N-k)} \ket{cb} \\
		&\quad+ \sqrt{2l(N-k)} \ket{cc} + (N-k) \ket{cd} \\
		&\quad+ \sqrt{k(N-k)} \ket{da} + (N-k) \ket{dc} \\
		&\quad+ \sqrt{2l(N-k)} \ket{dd} \Big].
\end{align*}
The evolution is depicted in \fref{fig:prob-1000-1000-5-5-symmetric}. Without self-loops, the success probability reaches $1$. With self-loops, however, the success probability is worse.

Let us prove this analytically. For large $N$, the initial state is asymptotically
\[ \ket{\psi(0)} = \frac{1}{\sqrt{2}}(\ket{cd}+\ket{dc}). \]
This can be written as a linear combination of asymptotic eigenvectors of $U$:
\[ \ket{\psi(0)} = a\ket{\psi_1}+b\ket{\psi_2}+c\ket{\psi_3}+d\ket{\psi_4}+e\ket{\psi_5}+f\ket{\psi_6}, \]
where
\begin{align*}
	& a=b=\frac{k}{2\sqrt{k(k+l)}}, \\
	& c=d=0, \\
	& e=\frac{-\sqrt{l}\sqrt{k+2l}}{\sqrt{2}(k+l)}, \\
	& f=\frac{-k}{\sqrt{2}(k+l)}.
\end{align*}
Then after $t$ steps, the system is in the state
\[ U^t \ket{\sigma} = a \lambda_1^t \ket{\psi_1} + b \lambda_2^t \ket{\psi_2} + e \lambda_5^t \ket{\psi_5} + f \lambda_6^t \ket{\psi_6}. \]
The success probability at time $t$ is
\begin{align*}
	p(t)
		&= |\bra{aa}U^t\ket{\sigma}|^2 + |\bra{ab}U^t\ket{\sigma}|^2 + |\bra{ad}U^t\ket{\sigma}|^2 \\
		&\quad+ |\bra{ba}U^t\ket{\sigma}|^2 + |\bra{bb}U^t\ket{\sigma}|^2 + |\bra{bc}U^t\ket{\sigma}|^2 \\
		&= \frac{k}{2(k+l)^2}[2k+3l-l\cos(\theta t)] \sin^2\left( \frac{\theta t}{2} \right).
\end{align*}
In order to find the runtime and maximum success probability of the system we take the first time derivative of $p(t)$,
\[ \frac{dp}{dt}=\frac{k\theta(k+2l-l\cos(\theta t))\sin(\theta t)}{2(k+l)^2}, \]
and set it equal to zero to yield a runtime of
\begin{equation}
	\label{eq:bothmarked_symmetric_runtime}
	t_* = \frac{\pi}{\theta} \approx \frac{\pi}{\sin \theta} = \frac{\pi}{2} \sqrt{\frac{N}{k+l}}.
\end{equation}
Substituting this runtime into $p(t)$ yields a maximum success probability of
\begin{equation}
	\label{eq:bothmarked_symmetric_prob}
	p_*=\frac{k(k+2l)}{(k+l)^2}. 
\end{equation}
These results are consistent with our numerical simulations in \fref{fig:prob-1000-1000-5-5-symmetric}. For example, the dashed red curve reaches its first peak at time $t_* = (\pi/2)\sqrt{2000/(10+5)} \approx 18$ with a height of $p_* = 10(10+2\cdot5)/(10 + 5)^2 \approx 0.889$. 

When the success probability is less than one, we may have to repeat the algorithm until a marked vertex is found. Including these repetitions, the total amount of time is
\[ T = \frac{t_*}{p_*} = \frac{\pi}{2} \frac{(k+l)^{3/2}}{k(k+2l)} \sqrt{N}. \]
This is minimized when $l = k/2$, at which the total runtime is
\[ \frac{3\pi}{8} \sqrt{\frac{3N}{2k}} \approx 1.443 \sqrt{\frac{N}{k}}, \]
which is an improvement over the loopless algorithm's $(\pi/2) \sqrt{N/k} \approx 1.571 \sqrt{N/k}$. This improvement, however, assumes the loopless algorithm evolves to its maximum success probability of $p_* = 1$. If we stop the loopless algorithm early, we can achieve a similar constant-factor speedup, as was explored for Grover's algorithm \cite{Zalka1999,Nahimovs2010}.


\subsection{General Problem}

\begin{figure}
\begin{center}
	\subfloat[] {
		\includegraphics[width=3in]{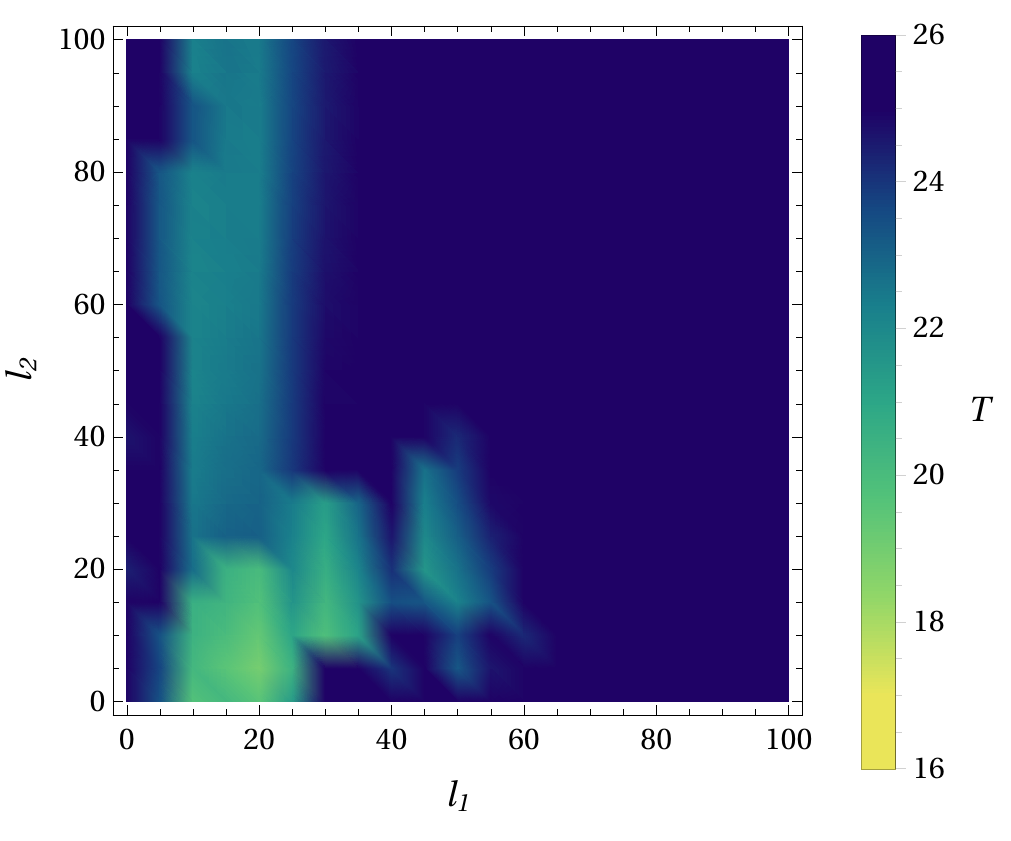}
                \label{fig:heatmap-s-500-1500-5-2}
        }

	\subfloat[] {
                \includegraphics[width=3in]{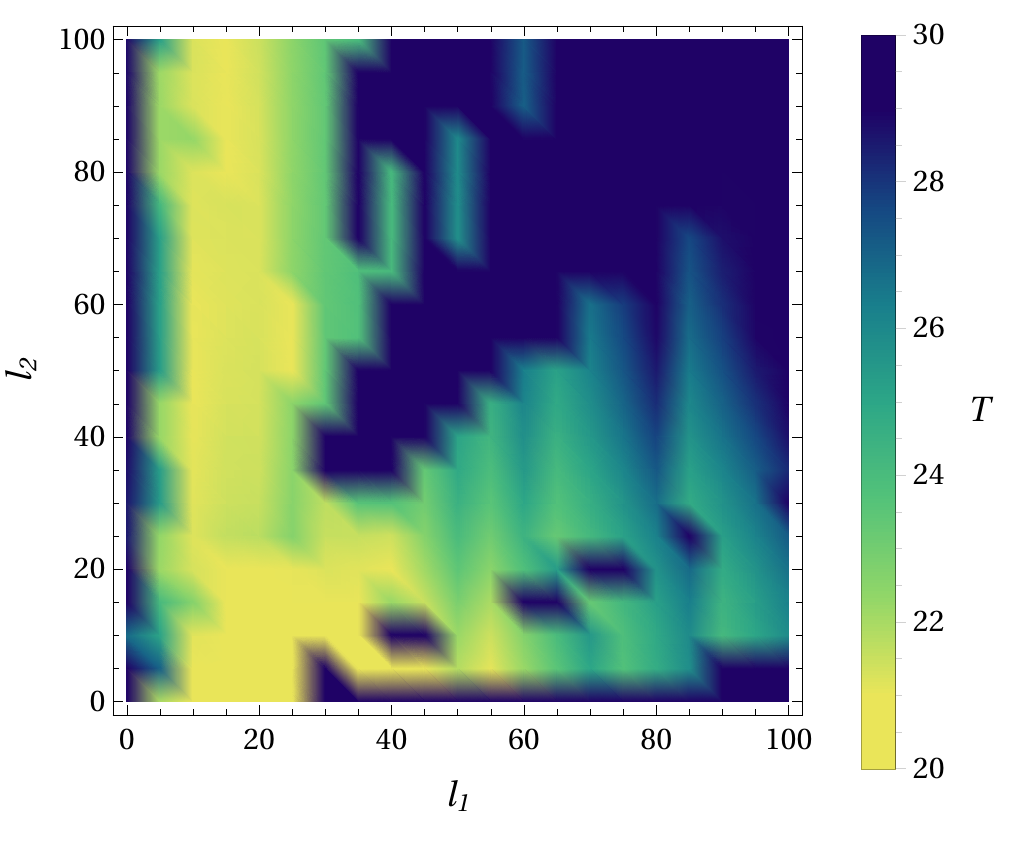}
                \label{fig:heatmap-sigma-500-1500-5-2}
        }
        
    \subfloat[] {
		\includegraphics{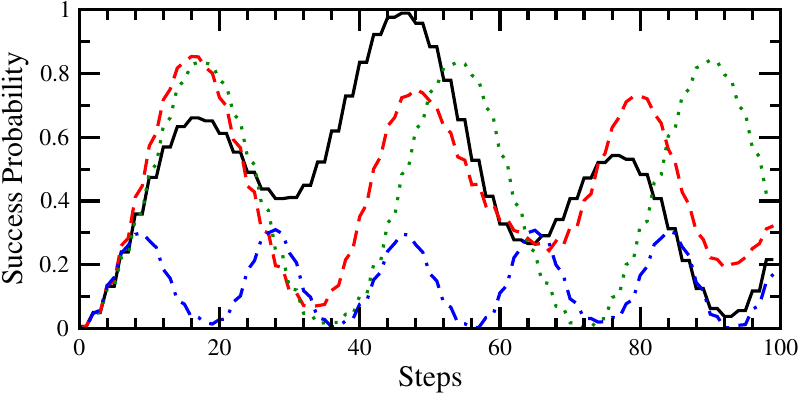}
                \label{fig:prob-sigma-500-1500-5-2}
    	}
	\caption{\label{fig:500-1500-5-2}Lackadaisical quantum search on the complete bipartite graph with $N_1=500$, $N_2=1500$, $k_1=5$, and $k_2=2$. The heatmaps in (a) and (b) depict the total runtime for initial states $\ket{s}$ and $\ket{\sigma}$, respectively, with various values of $l_1$ and $l_2$. The middle of the color scale corresponds to the loopless algorithm. In (c), the success probability as a function of time is shown for initial state $\ket{\sigma}$. The solid black curve is $l_1 = 0$, $l_2 = 0$, the dashed red curve is $l_1 = 15$, $l_2 = 5$, the dotted green curve is $l_1 = 15$, $l_2 = 100$, and the dot-dashed blue curve is $l_1 = l_2 = 80$. }
\end{center}
\end{figure}

\begin{figure}
\begin{center}
        \subfloat[] {
                \includegraphics[width=3in]{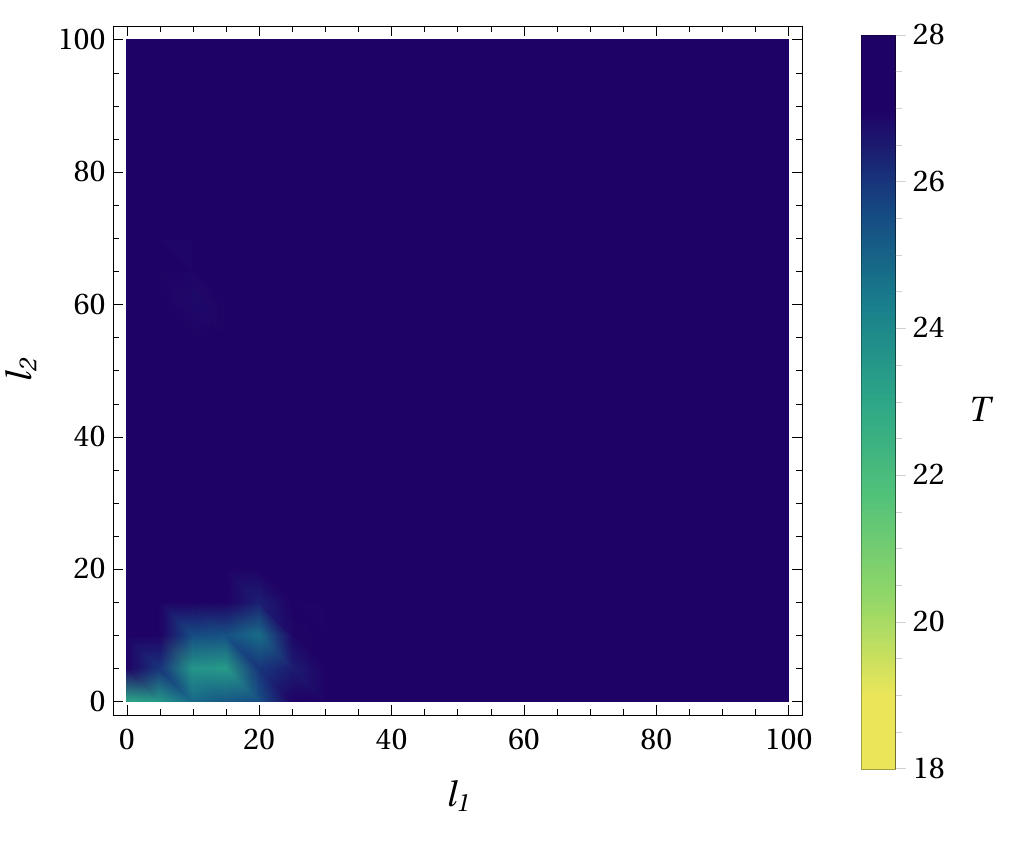}
                \label{fig:heatmap-s-500-1500-3-3}
        }

	\subfloat[] {
                \includegraphics[width=3in]{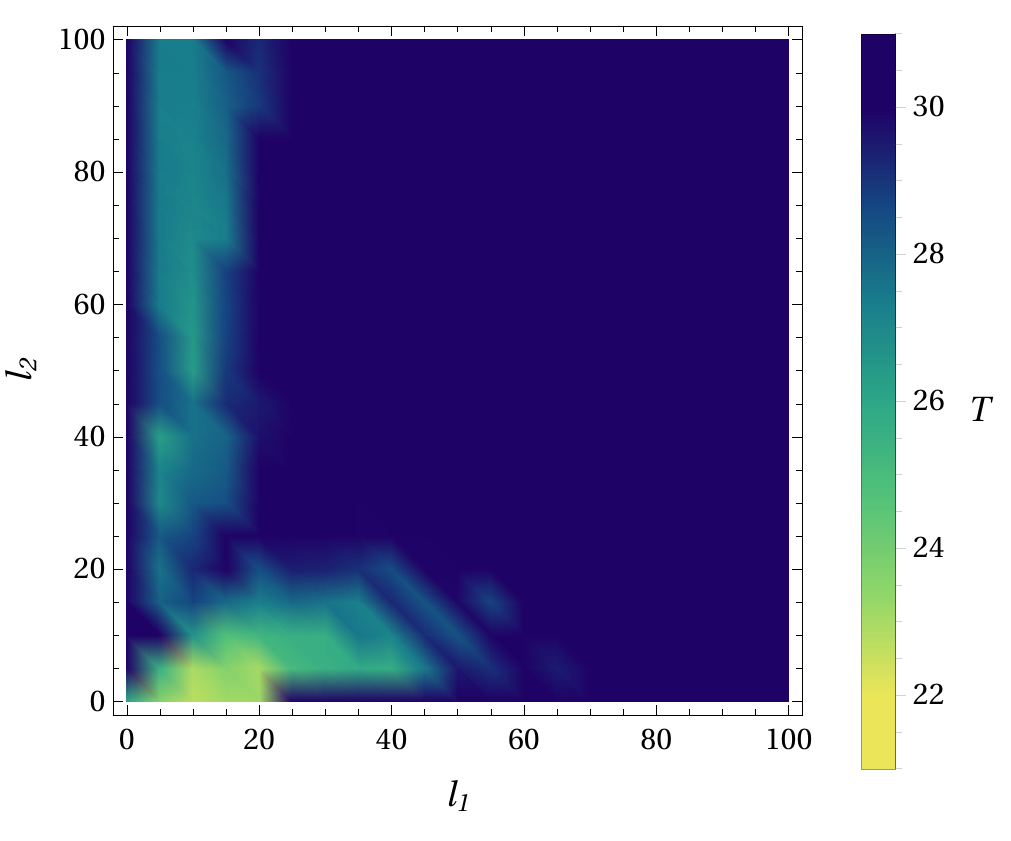}
                \label{fig:heatmap-sigma-500-1500-3-3}
        }

	\subfloat[] {
		\includegraphics{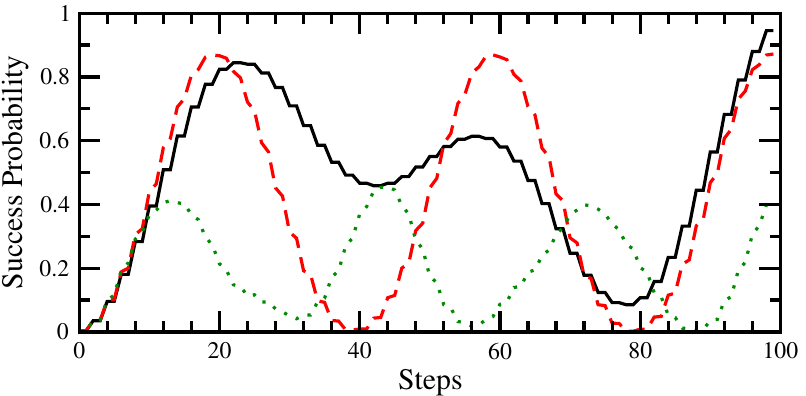}
                \label{fig:prob-sigma-500-1500-3-3}
        }
	\caption{Lackadaisical quantum search on the complete bipartite graph with $N_1=500$, $N_2=1500$, and $k_1 = k_2 = 3$. The heatmaps in (a) and (b) depict the total runtime for initial states $\ket{s}$ and $\ket{\sigma}$, respectively, with various values of $l_1$ and $l_2$. The middle of the color scale corresponds to the loopless algorithm. In (c), the success probability as a function of time is shown for initial state $\ket{\sigma}$. The solid black curve is $l_1 = 0$, $l_2 = 0$, the dashed red curve is $l_1 = 10$, $l_2 = 1$, and the dotted green curve is $l_1 = l_2 = 30$.}
\end{center}
\end{figure}

Now we numerically explore searching when the problem is not symmetric. As explained in the previous section, the success probability is not the only important quantity, since the time at which the success probability peaks is important as well. So for various values of $l_1$ and $l_2$, we numerically determine the total runtime with classical repetitions $T = t_* / p_*$, where $t_*$ is the time to the first maximum in success probability, and $p_*$ is the value of the first maximum in success probability. So smaller values of $T$ are better. Obtaining $T$ for a variety of values of $l_1$ and $l_2$, we plot $T$ as a heatmap in \fref{fig:heatmap-s-500-1500-5-2} with initial state $\ket{s}$ and in \fref{fig:heatmap-sigma-500-1500-5-2} with initial state $\ket{\sigma}$. In both of these plots, $N_1 = 500$, $N_2 = 1500$, $k_1 = 5$, $k_2 = 2$, and the loopless algorithm's value of $T$ corresponds to the center of the color scale. From \fref{fig:heatmap-s-500-1500-5-2}, self-loops do not appreciably improve search with initial state $\ket{s}$. But from \fref{fig:heatmap-sigma-500-1500-5-2}, the self-loops do improve the total runtime for many values of $l_1$ and $l_2$. As an example, in \fref{fig:prob-sigma-500-1500-5-2}, we plot the success probability vs time for several values of $l_1$ and $l_2$. The solid black curve corresponds to the loopless ($l_1  = l_2 = 0$) case, and although the success probability reaches 1 at the second peak, it is faster, on average, to measure the system at the first peak and risk repeating the algorithm. Our calculation of $T = t_*/p_*$ takes the first peak. The dashed red curve corresponds to $l_1 = 15$ and $l_2 = 5$, which has a faster total runtime than the loopless case. The dotted green curve corresponds to $l_1 = 15$ and $l_2 = 100$, and it evolves similarly to the red dashed curve with $l_1 = 15$ and $l_2 = 5$. Finally, the dot-dashed blue curve corresponds to $l_1 = l_2 = 80$, and this is slower than the loopless algorithm. Generally, weighted self-loops improve search from $\ket{\sigma}$ for a greater range of weights than from $\ket{s}$.

Next, in Figs.~\ref{fig:heatmap-s-500-1500-3-3} and \ref{fig:heatmap-sigma-500-1500-3-3}, we keep $N_1 = 500$ and $N_2 = 1500$, but now we take the number of marked vertices to be $k_1 = k_2 = 3$. In the former subfigure, the initial state is $\ket{s}$, and in the latter, it is $\ket{\sigma}$. As before, the center of the color scales correspond to the loopless algorithm's total runtime. We see some improvement in \fref{fig:heatmap-sigma-500-1500-3-3}, where the yellow area is. Exploring this in \fref{fig:prob-sigma-500-1500-3-3}, the solid back curve is the loopless algorithm and the dashed red curve is $l_1 = 10$, $l_2 = 1$, which is inside the region where we expect a speedup. Finally, the dotted green curve is $l_1 = l_2 = 30$, which is inside the region where we expect the algorithm to search poorly.

\begin{figure}
\begin{center}
        \subfloat[] {
                \includegraphics[width=3in]{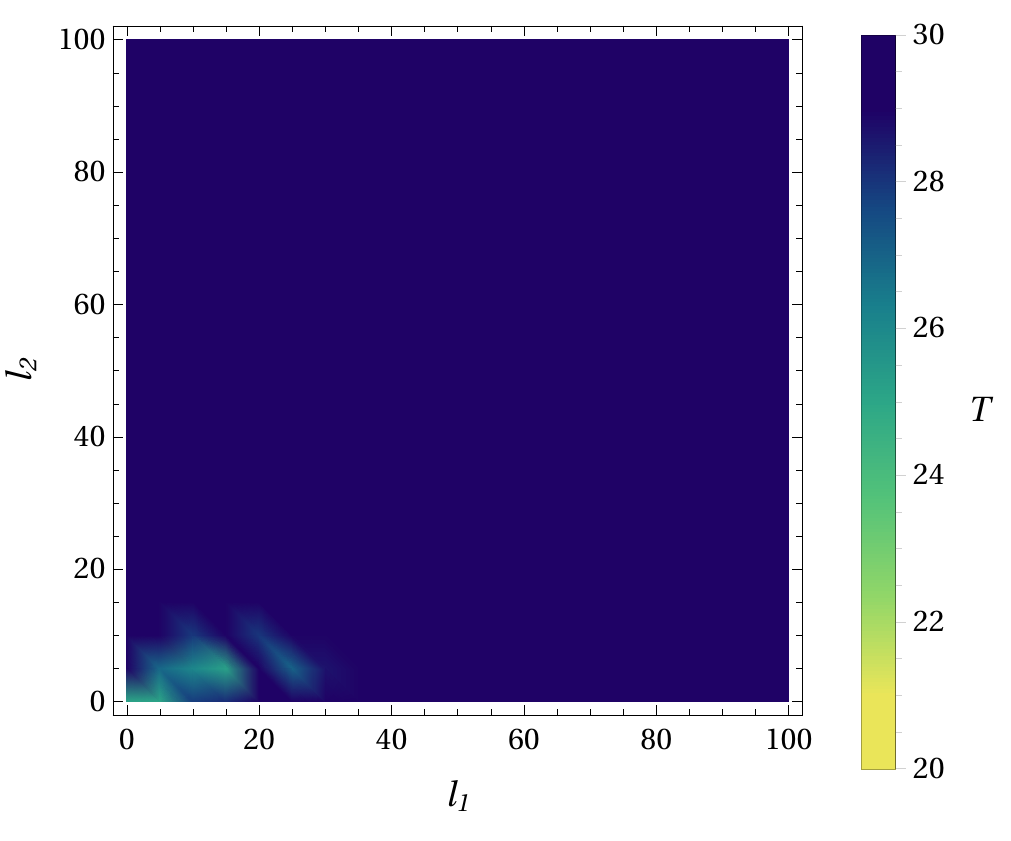}
                \label{fig:heatmap-s-500-1500-2-5}
        } \quad \quad
	\subfloat[] {
                \includegraphics[width=3in]{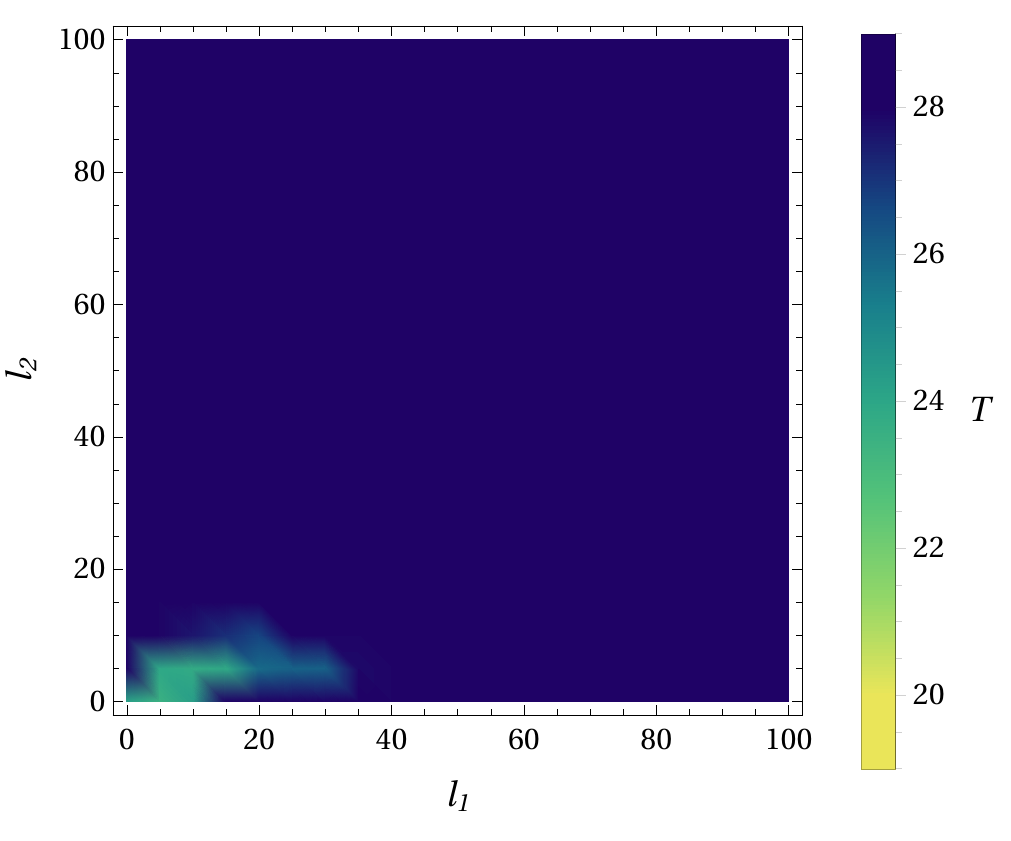}
                \label{fig:heatmap-sigma-500-1500-2-5}
        }  
	\caption{\label{fig:heatmap-500-1500-2-5}Heatmaps depicting the total runtime for the lackadaisical search on the complete bipartite graph with independently weighted self-loops in each set where $N_1=500$, $N_2=1500$, $k_1=2$, and $k_2=5$ for (a) initial state $\ket{s}$. (b) initial state $\ket{\sigma}$. The middle of the color scale corresponds to the loopless algorithm.}
\end{center}
\end{figure}

In \fref{fig:heatmap-500-1500-2-5}, we continue letting $N_1 = 500$ and $N_2 = 1500$, but now $k_1 = 2$, $k_2 = 5$ (this is opposite \fref{fig:500-1500-5-2}). For either initial state, there is no speedup. 

\begin{figure*}
\begin{center}
        \subfloat[] {
                \includegraphics[width=3in]{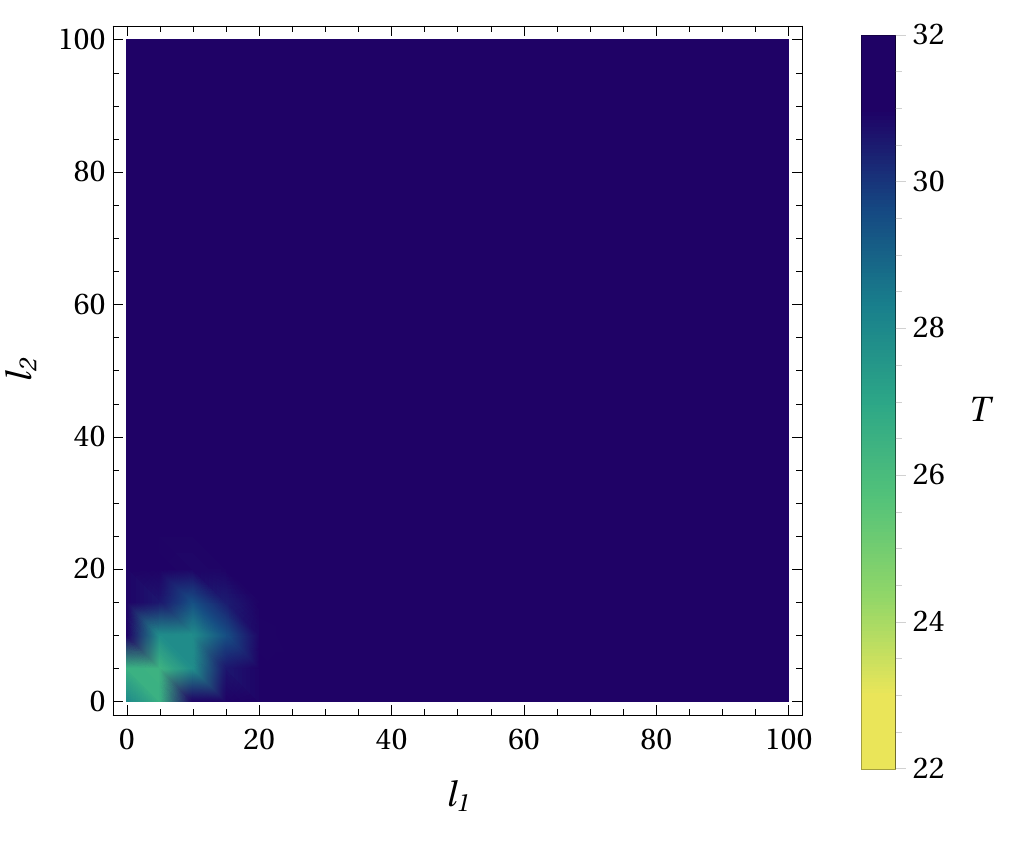}
                \label{fig:heatmap-s-1000-1000-3-3}
        } \quad \quad
        \subfloat[] {
                \includegraphics[width=3in]{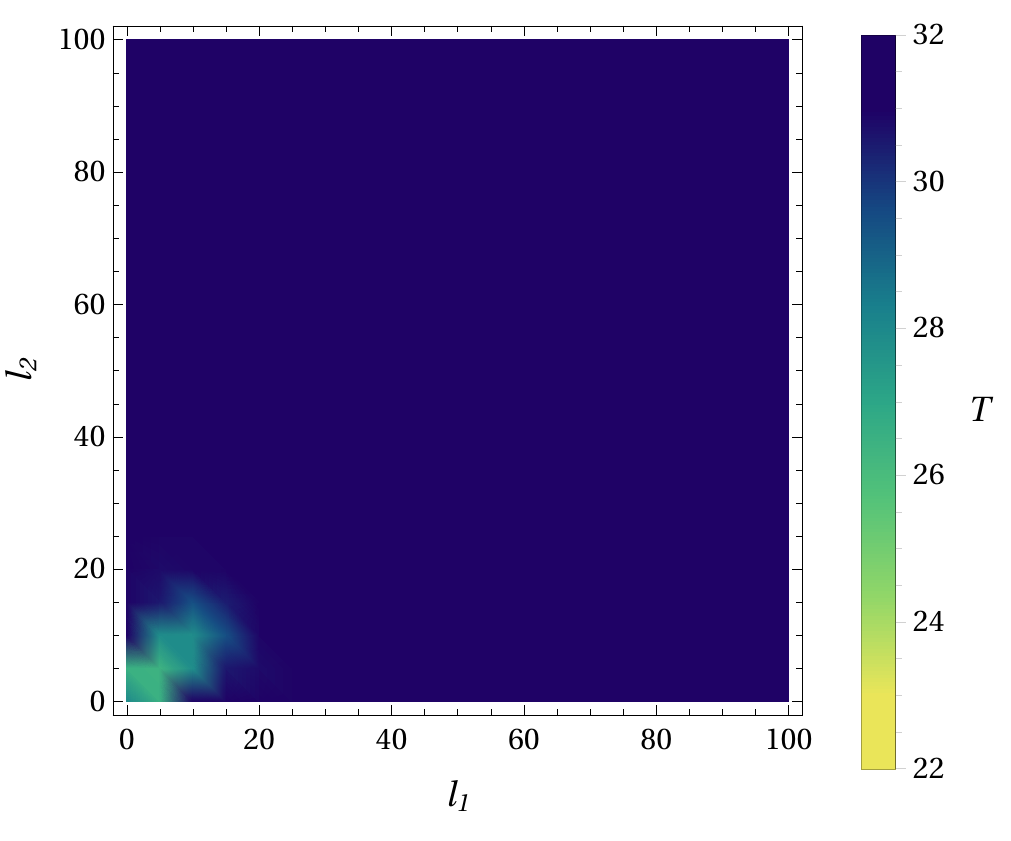}
                \label{fig:heatmap-sigma-1000-1000-3-3}
        }

        \subfloat[] {
                \includegraphics[width=3in]{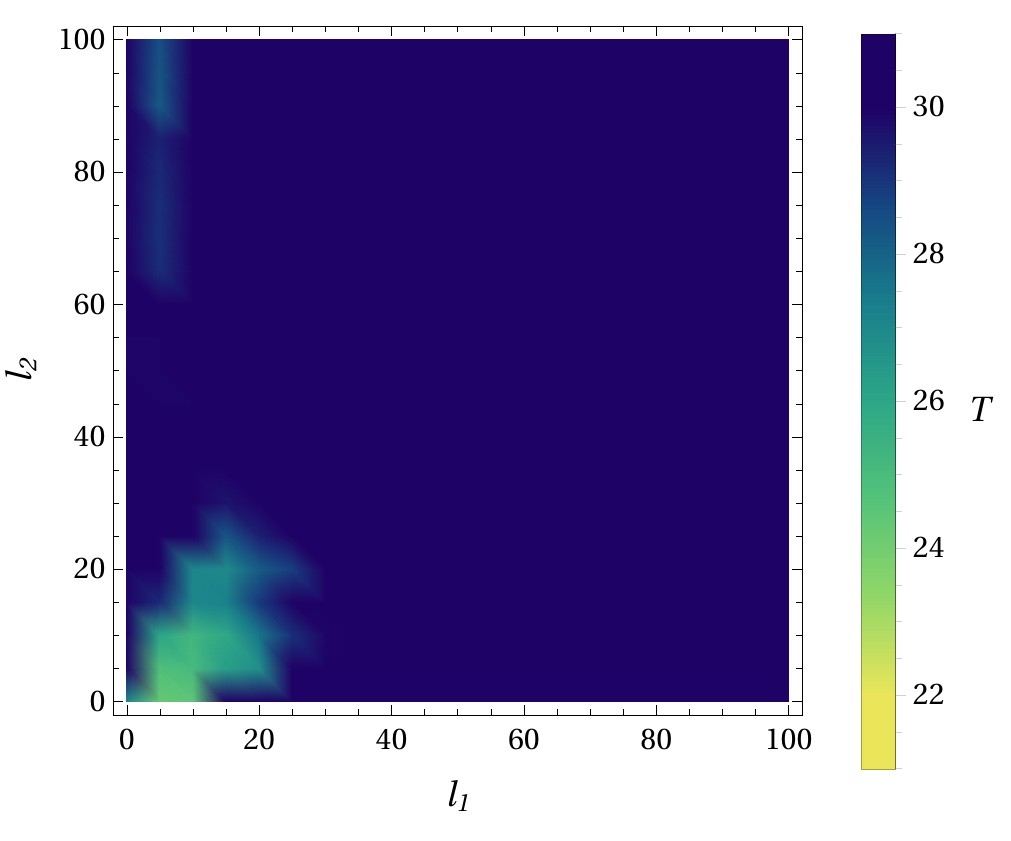}
                \label{fig:heatmap-s-1000-1000-5-2}
        } \quad\quad
        \subfloat[] {
                \includegraphics[width=3in]{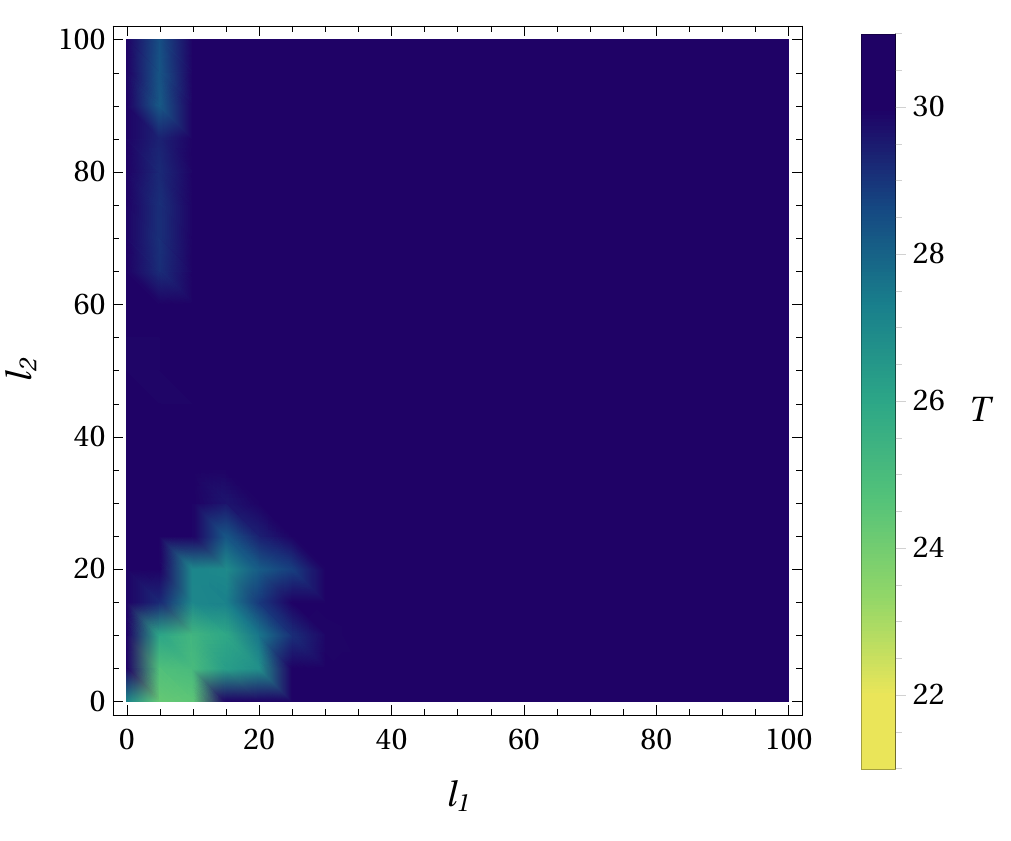}
                \label{fig:heatmap-sigma-1000-1000-5-2}
        }
	\caption{\label{fig:heatmap-1000-1000}Heatmaps depicting the total runtime for the lackadaisical search on the complete bipartite graph with independently weighted self-loops in each set where $N_1=N_2=1000$. In (a), $k_1=k_2=3$ for initial state $\ket{s}$. In (b), $k_1=k_2=3$ for initial state $\ket{\sigma}$. In (c), $k_1=5$ and $k_2=2$ for initial state $\ket{s}$. In (d), $k_1=5$ and $k_2=2$ for initial state $\ket{\sigma}$. The middle of the color scale corresponds to the loopless algorithm.}
\end{center}
\end{figure*}

Finally, in \fref{fig:heatmap-1000-1000}, we explore search when $N_1 = N_2 = 1000$. In Figs.~\ref{fig:heatmap-s-1000-1000-3-3} and \ref{fig:heatmap-sigma-1000-1000-3-3}, the number of marked vertices in each set is $k_1 = k_2 = 3$, but with respective initial states $\ket{s}$ and $\ket{\sigma}$. These figures are similar because $N_1$ and $N_2$ are large compared to $l_1$ and $l_2$, so the differences between the initial states are small. Note the heatmaps are identical, however, along the diagonal, since the diagonal corresponds to the symmetric problem that we analyzed in the previous section. From these heatmaps, there seems to be no improvement with self-loops. Similarly, in Figs.~\ref{fig:heatmap-s-1000-1000-5-2} and \ref{fig:heatmap-sigma-1000-1000-5-2}, we have $k_1 = 5$ and $k_2 = 2$, and while the heatmaps look similar, they are sightly different. Again, there seems to be no improvement over the loopless algorithm.


\section{Conclusion}

We have introduced the nonhomogeneous lackadaisical quantum walk, where self-loops at each vertex can have different weights. Such walks naturally arise on the complete bipartite graph, since vertices in each partite set have different structure, and hence their self-loops naturally carry different weights. We explored how nonhomogeneous lackadaisical quantum walks search the complete bipartite graph from two initial states: the uniform state $\ket{s}$ \eqref{eq:s} and the stationary state under the walk $\ket{\sigma}$ \eqref{eq:sigma}.

When the $k$ marked vertices are confined to a single partite set, say $X$, the nonhomogeneous lackadaisical quantum walk is faster with $\ket{s}$ when $N_2 > (3-2\sqrt{2})N_1$, i.e., when the graph is not too irregular. This takes $l_1 = kN_2/2N_1$\eqref{eq:onemarked_l1}, whereas $l_2$ can take any value (assuming $N_1$ and $N_2$ are large by comparison). On the other hand, when the initial state is $\ket{\sigma}$, then the nonhomogeneous lackadaisical quantum walk is always faster, no matter how irregular the graph may be, again when $l_1 = kN_2/2N_1$ and independent of $l_2$.

In the specific case where $k = 1$ and $N_1 = N_2$, loopless search on the complete bipartite graph and the complete graph behave the same way. With self-loops, however, the success probability is improved for the complete bipartite graph when $0 < l_1 < (3 + 2\sqrt{2}/2$, but it is improved for the complete graph when $0 < l < 3+2\sqrt{2}$. Hence, the range of weights for which we see an improvement when searching the complete bipartite graph is half that of the complete graph.

With marked vertices in both partite sets, we analytically proved that in the symmetric case where $N_1=N_2$, $k_1=k_2$, and $l_1=l_2$, the self-loops only worsened the search. For the various other cases that could not be proved analytically, simulations showed that speedups from self-loops only occur in special cases. Often, no speedup is obtained, no matter what values $l_1$ and $l_2$ take. This result is perhaps surprising given the success of lackadaisical quantum walks in prior research, and the freedom to choose $l_1$ and $l_2$ arbitrarily. This suggests that more research is required to understand when speedups are available using lackadaisical quantum walks.


\begin{acknowledgments}
	This work was partially supported by T.W.'s startup funds from Creighton University.
\end{acknowledgments}


\appendix
\section{\label{sec:eigen_onemarked}Eigensystem of the Search Operator with Marked Vertices in One Set}

The leading-order terms of $U$ \eqref{eq:onemarked_U} are
\[ U_0 = \begin{pmatrix}
	1 & 0 & 0 & 0 & 0 & 0 & 0 \\
	0 & 0 & -1 & 0 & 0 & 0 & 0 \\
	0 & -1 & 0 & 0 & 0 & 0 & 0 \\
	0 & 0 & 0 & -1 & 0 & 0 & 0 \\
	0 & 0 & 0 & 0 & 0 & 1 & 0 \\
	0 & 0 & 0 & 0 & 1 & 0 & 0 \\
	0 & 0 & 0 & 0 & 0 & 0 & -1 \\
\end{pmatrix}. \]
The normalized eigenvectors and respective eigenvalues of $U_0$ are
\begin{align*}
	\label{eq:lead_vecs}
	& \ket{v_1}=\frac{1}{\sqrt{2}}[0,0,0,0,1,1,0]^\intercal, \quad 1 \\
	& \ket{v_2}=\frac{1}{\sqrt{2}}[0,-1,1,0,0,0,0]^\intercal, \quad 1 \\
	& \ket{v_3}=[1,0,0,0,0,0,0]^\intercal, \quad 1 \\
	& \ket{v_4}=[0,0,0,0,0,0,1]^\intercal, \quad -1 \\
	& \ket{v_5}=\frac{1}{\sqrt{2}}[0,0,0,0,-1,1,0]^\intercal, \quad -1 \\
	& \ket{v_6}=[0,0,0,1,0,0,0]^\intercal, \quad -1 \\
	& \ket{v_7}=\frac{1}{\sqrt{2}}[0,1,1,0,0,0,0]^\intercal, \quad -1. \\
\end{align*}
Note the first three eigenvectors are degenerate with eigenvalue $1$, and the last four eigenvectors are degenerate with eigenvalue $-1$.

Now we include the perturbation. The next-leading-order terms of $U$ \eqref{eq:onemarked_U} are
\[ U_1= \begin{pmatrix}
	0 & -l_{12} & 0 & 0 & 0 & 0 & 0 \\
	0 & 0 & 0 & 0 & 2\sqrt{\frac{k}{N_1}} & 0 & 0 \\
	-l_{12} & 0 & 0 & 0 & 0 & 0 & 0 \\
	0 & 0 & 0 & 0 & l_{21} & 0 & 0 \\
	0 & 0 & 0 & 0 & 0 & 0 & l_{12} \\
	0 & 0 & 2\sqrt{\frac{k}{N_1}} & l_{21} & 0 & 0 & 0 \\
	0 & 0 & 0 & 0 & 0 & l_{12} & 0 \\
\end{pmatrix}, \]
where $l_{ij} = 2\sqrt{l_i/N_j}$. Since $\ket{v_1}$, $\ket{v_2}$, and $\ket{v_3}$ are degenerate eigenvectors of $U_0$, linear combinations $\alpha_1 \ket{v_1} + \alpha_2 \ket{v_2} + \alpha_3 \ket{v_3}$ are eigenvectors of $U_0 + U_1$. These coefficients satisfy the eigenvalue relation
\[ \begin{pmatrix}
	1 & \sqrt{\frac{k}{N_1}} & 0 \\
	-\sqrt{\frac{k}{N_1}} & 1 & -\sqrt{\frac{2l_1}{N_2}} \\
	0 & \sqrt{\frac{2l_1}{N_2}} & 1 \\
\end{pmatrix} \begin{pmatrix}
	\alpha_1 \\
	\alpha_2 \\
	\alpha_3
\end{pmatrix} = \lambda \begin{pmatrix}
	\alpha_1 \\
	\alpha_2 \\
	\alpha_3
\end{pmatrix}. \]
where the $(i,j)$th entry of the matrix is $\langle v_i | (U_0 + U_1) | v_j \rangle$. Solving this, the eigenvectors of $U_0 + U_1$ are asymptotically
\begin{widetext}
\begin{align*}
	& \ket{\psi_1}=\frac{1}{\sqrt{1+\frac{2l_1 N_1}{k N_2}}} \left( -\sqrt{\frac{2l_1 N_1}{k N_2}} \ket{v_1} + \ket{v_3} \right), \quad \lambda_1=1, \\
	& \ket{\psi_2}=\frac{1}{\sqrt{2+\frac{k N_2}{l_1 N_1}}} \left( \sqrt{\frac{k N_2}{2l_1 N_1}} \ket{v_1} - i\sqrt{\frac{2l_1 N_1+k N_2}{2l_1 N_1}} \ket{v_2} + \ket{v_3} \right), \quad \lambda_2=e^{-i\theta}, \\
	& \ket{\psi_3}=\frac{1}{\sqrt{2+\frac{k N_2}{l_1 N_1}}} \left( \sqrt{\frac{k N_2}{2l_1 N_1}} \ket{v_1} + i\sqrt{\frac{2l_1 N_1+k N_2}{2l_1 N_1}} \ket{v_2} + \ket{v_3} \right), \quad \lambda_3=e^{i\theta},
\end{align*}
\end{widetext}
where
\[ \sin \theta=\sqrt{\frac{2l_1 N_1+k N_2}{N_1N_2}}. \]
Substituting in for $\ket{v_1}$, $\ket{v_2}$, and $\ket{v_3}$ yields the eigenvectors in the main text.

Similarly, since $\ket{v_4}$, $\ket{v_5}$, $\ket{v_6}$, and $\ket{v_7}$ are degenerate eigenvectors of $U_0$, linear combinations $\alpha_4 \ket{v_4} + \alpha_5 \ket{v_5} + \alpha_6 \ket{v_6} + \alpha_7 \ket{7}$ are eigenvectors of $U_0 + U_1$. These coefficients satisfy the eigenvalue relation
\[ \begin{pmatrix}
	-1 & \sqrt{\frac{2l_1}{N_2}} & 0 & 0 \\
	-\sqrt{\frac{2 l_1}{N_2}} & -1 & \sqrt{\frac{2l_2}{N_1}} & \sqrt{\frac{k}{N_1}} \\
	0 & -\sqrt{\frac{2l_2}{N_1}} & -1 & 0 \\
       	0 & -\sqrt{\frac{k}{N_1}} & 0 & -1 \\
\end{pmatrix} \begin{pmatrix}
	\alpha_4 \\
	\alpha_5 \\
	\alpha_6 \\
	\alpha_7
\end{pmatrix} = \lambda \begin{pmatrix}
	\alpha_4 \\
	\alpha_5 \\
	\alpha_6 \\
	\alpha_7
\end{pmatrix}. \]
Solving this, the asymptotic eigenvectors and eigenvalues of $U_0 + U_1$ are
\begin{widetext}
\begin{align*}
	& \ket{\psi_4} = \frac{1}{\sqrt{1+\frac{k N_2}{2 l_1 N_1}}} \left( \sqrt{\frac{kN_2}{2l_1N_1}} \ket{v_4} + \ket{v_7} \right), \quad \lambda_4=-1, \\
	& \ket{\psi_5} = \frac{1}{\sqrt{1+\frac{l_2N_2}{l_1N_1}}} \left( \sqrt{\frac{l_2N_2}{l_1N_1}} \ket{v_4} + \ket{v_6} \right), \quad \lambda_5=-1, \\
	& \ket{\psi_6} = \frac{1}{\sqrt{2+\frac{4(l_1N_1+l_2N_2)}{k N_2}}} \left( -\sqrt{\frac{2l_1 N_1}{k N_2}} + \ket{v_4} + i\sqrt{\frac{2l_1N_1+(k+2l_2)N_2}{kN_2}} \ket{v_5} + \sqrt{\frac{2l_2}{k}} \ket{v_6} + \ket{v_7} \right), \quad \lambda_6=-e^{i\phi}, \\
	& \ket{\psi_7} = \frac{1}{\sqrt{2+\frac{4(l_1N_1+l_2N_2)}{k N_2}}} \left( -\sqrt{\frac{2l_1 N_1}{k N_2}} \ket{v_4} - i\sqrt{\frac{2l_1N_1+(k+2l_2)N_2}{kN_2}} \ket{v_5} + \sqrt{\frac{2l_2}{k}} \ket{v_6} + \ket{v_7} \right), \quad \lambda_7=-e^{-i\phi},
\end{align*}
\end{widetext}
where
\[ \sin\phi=\sqrt{\frac{2l_1N_1+(k+2l_2)N_2}{N_1N_2}}. \]
Note $\ket{\psi_4}$ and $\ket{\psi_5}$ are still degenerate with eigenvalue $-1$. Furthermore, they are not orthogonal, but they can easily be orthogonalized using the Gram-Schmidt procedure. For our problem, however, this is unnecessary since the projection of either starting state, $\ket{s}$ or $\ket{\sigma}$, onto the space spanned by $\ket{\psi_4}$ and $\ket{\psi_5}$ is zero. Finally, substituting $\ket{v_4}, \dots, \ket{v_7}$, we get the eigenvectors stated in the main text.


\section{\label{sec:eigen_bothmarked}Eigensystem of the Search Operator with Marked Vertices in Both Sets}

For large $N_1$ and $N_2$, the leading order terms of $U$ \eqref{eq:U_bothmarked} are
\setcounter{MaxMatrixCols}{12}
\[ U_0= \begin{pmatrix}
 	1 & 0 & 0 & 0 & 0 & 0 & 0 & 0 & 0 & 0 & 0 & 0 \\
	0 & 0 & 0 & 1 & 0 & 0 & 0 & 0 & 0 & 0 & 0 & 0 \\
	0 & 0 & 0 & 0 & 0 & 0 & 0 & 0 & 0 & -1 & 0 & 0 \\
	0 & 1 & 0 & 0 & 0 & 0 & 0 & 0 & 0 & 0 & 0 & 0 \\
	0 & 0 & 0 & 0 & 1 & 0 & 0 & 0 & 0 & 0 & 0 & 0 \\
	0 & 0 & 0 & 0 & 0 & 0 & -1 & 0 & 0 & 0 & 0 & 0 \\
	0 & 0 & 0 & 0 & 0 & -1 & 0 & 0 & 0 & 0 & 0 & 0 \\
	0 & 0 & 0 & 0 & 0 & 0 & 0 & -1 & 0 & 0 & 0 & 0 \\
       	0 & 0 & 0 & 0 & 0 & 0 & 0 & 0 & 0 & 0 & 1 & 0 \\
	0 & 0 & -1 & 0 & 0 & 0 & 0 & 0 & 0 & 0 & 0 & 0 \\
	0 & 0 & 0 & 0 & 0 & 0 & 0 & 0 & 1 & 0 & 0 & 0 \\
	0 & 0 & 0 & 0 & 0 & 0 & 0 & 0 & 0 & 0 & 0 & -1 
\end{pmatrix}. \]
This has the following normalized eigenvectors and eigenvalues:
\begin{align*}
	& \ket{v_1}=\frac{1}{\sqrt{2}}[0,0,0,0,0,0,0,0,1,0,1,0]^\intercal, \quad \lambda_1=1 \\
	& \ket{v_2}=\frac{1}{\sqrt{2}}[0,0,-1,0,0,0,0,0,0,1,0,0]^\intercal, \quad \lambda_2=1 \\
	& \ket{v_3}=\frac{1}{\sqrt{2}}[0,0,0,0,0,-1,1,0,0,0,0,0]^\intercal, \quad \lambda_3=1 \\
	& \ket{v_4}=[0,0,0,0,1,0,0,0,0,0,0,0]^\intercal, \quad \lambda_4=1 \\
	& \ket{v_5}=\frac{1}{\sqrt{2}}[0,1,0,1,0,0,0,0,0,0,0,0]^\intercal, \quad \lambda_5=1 \\
	& \ket{v_6}=[1,0,0,0,0,0,0,0,0,0,0,0]^\intercal, \quad \lambda_6=1 \\
	& \ket{v_7}=[0,0,0,0,0,0,0,0,0,0,0,1]^\intercal, \quad \lambda_7=-1 \\
	& \ket{v_8}=\frac{1}{\sqrt{2}}[0,0,0,0,0,0,0,0,-1,0,1,0]^\intercal, \quad \lambda_8=-1 \\
	& \ket{v_9}=\frac{1}{\sqrt{2}}[0,0,1,0,0,0,0,0,0,1,0,0]^\intercal, \quad \lambda_9=-1 \\
	& \ket{v_{10}}=[0,0,0,0,0,0,0,1,0,0,0,0]^\intercal, \quad \lambda_{10}=-1 \\
	& \ket{v_{11}}=\frac{1}{\sqrt{2}}[0,0,0,0,0,1,1,0,0,0,0,0]^\intercal, \quad \lambda_{11}=-1 \\
	& \ket{v_{12}}=\frac{1}{\sqrt{2}}[0,-1,0,1,0,0,0,0,0,0,0,0]^\intercal, \quad \lambda_{12}=-1
\label{eq:vecs2}
\end{align*}
Note the first six eigenvectors are degenerate with eigenvalue $1$, and the last six eigenvectors are degenerate with eigenvalue $-1$.

Now including the perturbation, the next-leading terms of $U$ are
\begin{widetext}
\[ U_1= \begin{pmatrix} 
 	0 & 0 & \frac{-2\sqrt{l_1}}{\sqrt{N_2}} & 0 & 0 & 0 & 0 & 0 & 0 & 0 & 0 & 0 \\
	0 & 0 & 0 & 0 & 0 & \frac{-2\sqrt{k_1}}{\sqrt{N_1}} & 0 & 0 & 0 & 0 & 0 & 0 \\
       	0 & 0 & 0 & 0 & 0 & 0 & 0 & 0 & 0 & 0 & \frac{2\sqrt{k_1}}{\sqrt{N_1}} & 0 \\
	0 & 0 & \frac{-2\sqrt{k_2}}{\sqrt{N_2}} & 0 & 0 & 0 & 0 & 0 & 0 & 0 & 0 & 0 \\
	0 & 0 & 0 & 0 & 0 & \frac{-2\sqrt{l_2}}{\sqrt{N_1}} & 0 & 0 & 0 & 0 & 0 & 0 \\
	0 & 0 & 0 & 0 & 0 & 0 & 0 & 0 & \frac{2\sqrt{k_2}}{\sqrt{N_2}} & 0 & 0 & 0 \\
	0 & 0 & 0 & \frac{-2\sqrt{k_1}}{\sqrt{N_1}} & \frac{-2\sqrt{l_2}}{\sqrt{N_1}} & 0 & 0 & 0 & 0 & 0 & 0 & 0 \\
	0 & 0 & 0 & 0 & 0 & 0 & 0 & 0 & \frac{2\sqrt{l_1}}{\sqrt{N_2}} & 0 & 0 & 0 \\
      	0 & 0 & 0 & 0 & 0 & 0 & 0 & 0 & 0 & \frac{2\sqrt{k_1}}{\sqrt{N_1}} & 0 & \frac{2\sqrt{l_2}}{\sqrt{N_1}} \\
	\frac{-2\sqrt{l_1}}{\sqrt{N_2}} & \frac{-2\sqrt{k_2}}{\sqrt{N_2}} & 0 & 0 & 0 & 0 & 0 & 0 & 0 & 0 & 0 & 0 \\
	0 & 0 & 0 & 0 & 0 & 0 & \frac{2\sqrt{k_2}}{\sqrt{N_2}} & \frac{2\sqrt{l_1}}{\sqrt{N_2}} & 0 & 0 & 0 & 0 \\
	0 & 0 & 0 & 0 & 0 & 0 & 0 & 0 & 0 & 0 & \frac{2\sqrt{l_2}}{\sqrt{N_1}} & 0 
\end{pmatrix}. \]
Since $\ket{v_1}, \dots, \ket{v_6}$ are degenerate eigenvectors of $U_0$, linear combinations $\alpha_1 \ket{v_1} + \alpha_2 \ket{v_2} + \dots + \alpha_6 \ket{v_6}$ are eigenvectors of $U_0 + U_1$. These coefficients satisfy
\[ \begin{pmatrix}
 	1 & \sqrt{\frac{k_1}{N_1}} & \sqrt{\frac{k_2}{N_2}} & 0 & 0 & 0 \\
	-\sqrt{\frac{k_1}{N_1}} & 1 & 0 & 0 & -\sqrt{\frac{k_2}{N_2}} & -\sqrt{\frac{2l_1}{N_2}} \\
	-\sqrt{\frac{k_2}{N_2}} & 0 & 1 & -\sqrt{\frac{2l_2}{N_1}} & -\sqrt{\frac{k_1}{N_1}} & 0 \\
	0 & 0 & \sqrt{\frac{2l_2}{N_1}} & 1 & 0 & 0 \\
	0 & \sqrt{\frac{k_2}{N_2}} & \sqrt{\frac{k_1}{N_1}} & 0 & 1 & 0 \\
	0 & \sqrt{\frac{2l_1}{N_2}} & 0 & 0 & 0 & 1 
\end{pmatrix} \begin{pmatrix}
	\alpha_1 \\
	\alpha_2 \\
	\alpha_3 \\
	\alpha_4 \\
	\alpha_5 \\
	\alpha_6 \\
\end{pmatrix} = E \begin{pmatrix}
	\alpha_1 \\
	\alpha_2 \\
	\alpha_3 \\
	\alpha_4 \\
	\alpha_5 \\
	\alpha_6 \\
\end{pmatrix}, \]
where the entries of the matrices are $\langle v_i | (U_0 + U_1) | v_j \rangle$. Solving this eigenvalue relation is possible, but the solution is sufficiently messy such that interpreting the result is difficult. So instead, we solve it assuming $N_1 = N_2 = N/2$, $l_1 = l_2 = l$, and $k_1 = k_2 = k/2$ . Then the asymptotic eigenvectors of $U_0 + U_1$ are
\begin{align*}
	& \ket{\psi_1} = \frac{1}{\sqrt{4+\frac{4k}{l}}} \left( \sqrt{\frac{k}{l}} \ket{v_1} - \frac{i\sqrt{k+l}}{\sqrt{l}} \ket{v_2} - \frac{i\sqrt{k+l}}{\sqrt{l}} \ket{v_3} + \ket{v_4} + \sqrt{\frac{k}{l}} \ket{v_5} + \ket{v_6} \right), \quad \lambda_1=e^{-i\theta}, \\
	& \ket{\psi_2} = \frac{1}{\sqrt{4+\frac{4k}{l}}} \left( \sqrt{\frac{k}{l}} \ket{v_1} + \frac{i\sqrt{k+l}}{\sqrt{l}} \ket{v_2} + \frac{i\sqrt{k+l}}{\sqrt{l}} \ket{v_3} + \ket{v_4} + \sqrt{\frac{k}{l}} \ket{v_5} + \ket{v_6} \right), \quad \lambda_2=e^{i\theta}, \\
	& \ket{\psi_3} = \frac{1}{2} \left( -i \ket{v_2} + i \ket{v_3} - \ket{v_4} + \ket{v_6} \right), \quad \lambda_3=e^{-i\phi}, \\
	& \ket{\psi_4} = \frac{1}{2} \left( i \ket{v_2} - i \ket{v_3} - \ket{v_4} + \ket{v_6} \right), \quad \lambda_3=e^{i\phi}, \\
	& \ket{\psi_5} = \frac{1}{\sqrt{2+\frac{4l}{k}}} \left( -2\sqrt{\frac{l}{k}} \ket{v_1} + \ket{v_4} + \ket{v_6} \right), \quad \lambda_5=1, \\
	& \ket{\psi_6} = \frac{1}{\sqrt{2}} \left( -\ket{v_1} + \ket{v_5} \right), \quad \lambda_6=1,
\end{align*}
where $\sin\theta = 2\sqrt{k+l}/\sqrt{N}$ and $\sin\phi = 2\sqrt{l}/\sqrt{N}$. Note $\ket{\psi_5}$ and $\ket{\psi_6}$ are still degenerate, and they are not orthogonal. We can orthogonalize them using the Gram-Schmidt procedure, but as shown in the main text, this is not necessary since both initial states $\ket{s}$ and $\ket{\sigma}$ can be expressed as a linear combination of $\ket{\psi_1}, \dots, \ket{\psi_6}$ as written. Substituting for $\ket{v_1}, \dots, \ket{v_6}$ yields the eigenvectors stated in the main text.

For the remaining eigenvectors, $\ket{v_7}, \dots, \ket{v_{12}}$ are degenerate eigenvectors of $U_0$, linear combinations $\alpha_7 \ket{v_7} + \alpha_8 \ket{v_8} + \dots + \alpha_{12} \ket{v_{12}}$ are also eigenvectors of $U_0 + U_1$, where the coefficients satisfy
\[ \begin{pmatrix}
 	-1 & \sqrt{\frac{2l_2}{N_1}} & 0 & 0 & 0 & 0 \\
	-\sqrt{\frac{2l_2}{N_1}} & -1 & -\sqrt{\frac{k_1}{N_1}} & \sqrt{\frac{2l_1}{N_2}} & \sqrt{\frac{k_2}{N_2}} & 0 \\
	0 & \sqrt{\frac{k_1}{N_1}} & -1 & 0 & 0 & \sqrt{\frac{k_2}{N_2}} \\
	0 & -\sqrt{\frac{2l_1}{N_2}} & 0 & -1 & 0 & 0 \\
	0 & -\sqrt{\frac{k_2}{N_2}} & 0 & 0 & -1 & -\sqrt{\frac{k_1}{N_1}} \\
	0 & 0 & -\sqrt{\frac{k_2}{N_2}} & 0 & \sqrt{\frac{k_1}{N_1}} & -1 
\end{pmatrix} \begin{pmatrix}
	\alpha_7 \\
	\alpha_8 \\
	\alpha_9 \\
	\alpha_{10} \\
	\alpha_{11} \\
	\alpha_{12} \\
\end{pmatrix} = E \begin{pmatrix}
	\alpha_7 \\
	\alpha_8 \\
	\alpha_9 \\
	\alpha_{10} \\
	\alpha_{11} \\
	\alpha_{12} \\
\end{pmatrix}. \]
\end{widetext}
Again, solving this is possible, but the resulting eigenvectors are too messy to interpret. We could solve them assuming $N_1 = N_2 = N/2$, $l_1 = l_2 = l$, and $k_1 = k_2 = k/2$, but this is unnecessary, since in this case, it is possible to express the initial states $\ket{s}$ and $\ket{\sigma}$ in terms of $\ket{\psi_1}, \dots, \ket{\psi_6}$ alone.


\bibliography{refs}

\end{document}